
\input phyzzx.tex
\input tables.tex

\def\rbtau{R_{b/\tau}}
\def\bmu{B^0_\mu}
\def\bz{B^0_Z}
\def\delgs{\delta_{GS}}
\def\sdil{SD}
\def\sns{SMS}
\def\ns{MS}
\def\dil{D}
\def\sdilp{\sdil^+}
\def\sdilm{\sdil^-}
\def\nsp{\ns^+}
\def\nsm{\ns^-}
\def\dilp{\dil^+}
\def\dilm{\dil^-}
\def\snsp{\sns^+}
\def\snsm{\sns^-}


\def\dilpi{\dilp_4}
\def\dilpii{\dilp_2}
\def\dilpiii{\dilp_7}
\def\dilpiv{\dilp_8}
\def\dilpv{\dilp_5}
\def\dilpvi{\dilp_3}
\def\dilpvii{\dilp_1}
\def\dilpviii{\dilp_6}

\def\dilmi{\dilm_6}
\def\dilmii{\dilm_3}
\def\dilmiii{\dilm_8}
\def\dilmiv{\dilm_9}
\def\dilmv{\dilm_7}
\def\dilmvi{\dilm_5}
\def\dilmvii{\dilm_1}
\def\dilmviii{\dilm_2}
\def\dilmix{\dilm_4}

\def\sdilpiii{\sdilp_1}
\def\sdilpvii{\sdilp_2}
\def\sdilpviii{\sdilp_3}

\def\sdilmv{\sdilm_1}
\def\sdilmviii{\sdilm_2}
\def\sdilmix{\sdilm_3}

\def\nspi{\nsp_1}
\def\nspii{\nsp_2}
\def\nspiii{\nsp_7}
\def\nspiv{\nsp_8}
\def\nspv{\nsp_4}
\def\nspvi{\nsp_6}
\def\nspvii{\nsp_5}
\def\nspviii{\nsp_3}

\def\nsmi{\nsm_1}
\def\nsmii{\nsm_2}
\def\nsmiii{\nsm_7}
\def\nsmiv{\nsm_3}
\def\nsmv{\nsm_4}
\def\nsmvi{\nsm_5}
\def\nsmvii{\nsm_6}

\def\snspvii{\snsp_2}
\def\snspviii{\snsp_3}
\def\snspix{\snsp_1}

\def\snsmii{\snsm_7}
\def\snsmiii{\snsm_5}
\def\snsmxi{\snsm_4}
\def\snsmxii{\snsm_1}
\def\snsmxiii{\snsm_2}
\def\snsmxiv{\snsm_3}

\def\ell{l}
\def\ep{e^+}
\def\em{e^-}
\def\mgrtino{m_{3/2}}
\def\mgltb{$\mgl\,$--$\tanb$}
\def\hl{h^0}

\def\ha{A^0}

\def\mhl{m_{\hl}}

\def\mha{m_{\ha}}

\def\gev{~{\rm GeV}}
\def\tev{~{\rm TeV}}
\def\pbi{~{\rm pb}^{-1}}
\def\fbi{~{\rm fb}^{-1}}


\def\prdj#1{{\it Phys. Rev.} {\bf D{#1}}}
\def\npbj#1{{\it Nucl. Phys.} {\bf B{#1}}}
\def\prlj#1{{\it Phys. Rev. Lett.} {\bf {#1}}}
\def\plbj#1{{\it Phys. Lett.} {\bf B{#1}}}

\def\prepj#1{{\it Phys. Rep.} {\bf {#1}}}

\def\mt{m_t}
\def\wp{W^+}
\def\wm{W^-}
\def\wpm{W^{\pm}}

\def\rta{\rightarrow}

\def\prdj#1{{\it Phys. Rev.} {\bf D{#1}}}
\def\npbj#1{{\it Nucl. Phys.} {\bf B{#1}}}
\def\prlj#1{{\it Phys. Rev. Lett.} {\bf {#1}}}
\def\plbj#1{{\it Phys. Lett.} {\bf B{#1}}}

\def\prepj#1{{\it Phys. Rep.} {\bf {#1}}}

\def\anti{\overline}
\def\pbi{~{\rm pb}^{-1}}
\def\fbi{~{\rm fb}^{-1}}

\catcode`\@=11 

\def\t1{{\tilde 1}}

\def\slash#1{#1\hskip-6pt/\hskip2pt}

\def\etmiss{\slash E_T}
\def\etmissvec{\vec \slash E_T}

\def\gev{\,{\rm GeV}}
\def\tev{\,{\rm TeV}}

\def\wt{\widetilde}

\def\rta{\rightarrow}

\def\gl{\wt g}
\def\mgl{m_{\gl}}
\def\stop{\wt t}
\def\stopi{\wt t_1}

\def\mstopi{m_{\stopi}}
\def\sq{\wt q}

\def\msq{m_{\sq}}
\def\slep{\wt \ell}

\def\slepl{\slep_L}
\def\slepr{\slep_R}
\def\mslepl{m_{\slepl}}
\def\mslepr{m_{\slepr}}
\def\sbot{\wt b}

\def\hl{h^0}

\def\ha{A^0}

\def\mhl{m_{\hl}}

\def\mha{m_{\ha}}

\def\tanb{\tan\beta}
\def\mt{m_t}
\def\mb{m_b}
\def\mz{m_Z}
\def\mgut{M_U}
\def\mstring{M_S}
\def\wp{W^+}
\def\wm{W^-}
\def\wpm{W^{\pm}}

\def\cnone{\wt\chi^0_1}
\def\cntwo{\wt\chi^0_2}
\def\snu{\wt\nu}

\def\msnu{m_{\snu}}
\def\mcnone{m_{\cnone}}
\def\mcntwo{m_{\cntwo}}

\def\chitil{\wt \chi}
\def\cpone{\wt \chi^+_1}
\def\cmone{\wt \chi^-_1}
\def\cpmone{\wt \chi^{\pm}_1}
\def\cmpone{\wt \chi^{\mp}_1}
\def\mcpone{m_{\cpone}}
\def\mcpmone{m_{\cpmone}}
\def\stauone{\wt \tau_1}
\def\stau{\wt \tau}
\def\mstauone{m_{\stauone}}
\def\tanb{\tan\beta}

\def\mz{m_Z}
\def\anti{\overline}

\def\ifmath#1{\relax\ifmmode #1\else $#1$\fi}

\def\3quarter{{\textstyle{3 \over 4}}}

\input phyzzx
\Pubnum={$\caps UCD-94-19$\cr $\caps FSU-HEP-940501$ \cr}
\date{May, 1994}

\titlepage
\vskip 0.75in
\baselineskip 0pt
\hsize=6.5in
\vsize=8.5in
\centerline{{\bf Tevatron and LEP-II Probes of }}
\vskip .075in
\centerline{{\bf Minimal and
String-Motivated Supergravity Models}}
\vskip .075in
\centerline{H. Baer$^a$, J.F. Gunion$^b$, C. Kao$^a$ and H. Pois$^b$}
\vskip .075in
\centerline{\it a) Dept. of Physics, Florida State University, Tallahassee,
FL 32306}
\centerline{\it b) Davis Institute for High Energy Physics,
Dept. of Physics, U.C. Davis, Davis, CA 95616}

\vskip .075in
\centerline{\bf Abstract}
\vskip .075in
\centerline{\Tenpoint\baselineskip=12pt
\vbox{\hsize=12.4cm
\noindent
We explore the ability of the Tevatron to probe Minimal Supersymmetry
with high energy scale boundary conditions
motivated by supersymmetry breaking in
the context of supergravity/superstring theory. A number of boundary
condition possibilities are considered: dilaton-like string boundary
conditions applied at the standard GUT unification scale or alternatively
at the string scale; and extreme (``no-scale'') minimal supergravity
boundary conditions imposed at the GUT scale or string scale.
For numerous specific cases within each scenario
the sparticle spectra are computed and then fed into ISAJET 7.07 so that
explicit signatures can be examined in detail. We find
that, for some of the boundary condition choices, large regions of parameter
space can be explored via same-sign dilepton
and isolated trilepton signals. For other choices, the mass reach of Tevatron
collider experiments is much more limited. We also compare mass reach of
Tevatron experiments with the corresponding reach at LEP 200.
}}

\vskip .15in
\noindent{\bf 1. Introduction}
\vskip .075in

Assessing our ability to experimentally probe
supersymmetric extensions of the Standard Model (SM)
at existing and future accelerators is a crucial issue for the
future of high energy physics.  Indeed,
$N=1$ supersymmetric models containing Standard Model matter and
gauge fields (and their superpartners) along with exactly two Higgs doublets
are remarkable in that the observed values of $\alpha_{QED}$,
$\sin^2\theta_W$ and $\alpha_s$ at the scale $\mz$ are highly
consistent with unification of the $U(1)$, $SU(2)_L$ and $SU(3)$ gauge
coupling constants at a scale of order $\mgut\sim 2\times 10^{16}\gev$.
\Ref\langacker{P. Langacker and M. Luo, \prdj{44} (1991) 817;
U. Amaldi, W. de Boer and H. Furstenau, \plbj{260} (1991) 447;
J. Ellis, S. Kelley and D. Nanopoulos, \plbj{260} (1991) 131.}
(Although additional singlet Higgs fields do not affect the unification,
we shall focus here
on the Minimal Supersymmetric Model (MSSM) in which there are only
two Higgs doublet fields, and no extra Higgs singlet field(s).)
In a completely general context, the
large uncertainty in the soft-supersymmetry-breaking
parameters of the MSSM makes it difficult to arrive at
definite predictions for the best probes and ultimate experimental
accessibility of the superparticles.  Even
the basic superpartner mass scales (which we
generically denote by $M_{SUSY}$) are rather uncertain, although
it is widely accepted that
they should lie below about $1\tev$ in order to provide an
obvious solution to the naturalness problem for the Higgs mass, and, in
addition the gauge-coupling
unification is only successful if $M_{SUSY}\lsim 1\tev$.
However, the success of gauge-coupling unification suggests that we should
consider models that also have relatively simple and universal boundary
conditions for the soft-supersymmetry-breaking parameters
at the unification scale. Implications at low-energies ($\lsim 1\tev$)
can then be obtained by renormalization group evolution
of the high-energy-scale parameters.

Supergravity and superstring theories provide the most attractive context
in which gauge unification based on minimal $N=1$ supersymmetry
can be natural. Superstring theory stands out as the only
candidate which is known to lead
to a consistent theory of quantum gravity. Physics below the Planck
scale is determined by the effective non-renormalizable supergravity
theory that is believed to arise from the string once the
super-heavy $M_P$-scale fields are integrated out.
Our goal in this paper will be to assess the extent
to which the Tevatron and LEP-II can probe the
superparticle spectrum of the MSSM with soft supersymmetry breaking
specified by boundary conditions (at the unification scale)
as predicted in a limited, but very attractive,
set of string and minimal supergravity models. The main focus
of the paper will be on assessing a wide range of possible
signals at the Tevatron,
including the missing-transverse-momentum, same-sign-lepton,
tri-lepton, and four-lepton discovery modes. We shall contrast
the parameter space range for which the Tevatron
can probe the superparticle spectrum with that for which
supersymmetry can be observed at LEP-II via
chargino-pair, slepton-pair and/or $Z+$Higgs associated production.

The outline of the paper follows.  In Sec. 2, we discuss
the motivation behind and nature of the string theory ``dilaton-like'' and
``no-scale'' minimal supergravity boundary conditions that we employ.
In Sec. 3, we delineate the (two-dimensional)
parameter spaces that are allowed
for each of the eight resulting models, given existing theoretical
and experimental constraints and our assumed top-quark mass of
$\mt(\mt)=170\gev$.  The all-important mass spectra and the consequent
general phenomenological implications are also detailed in Sec. 3.
In Sec. 4, we specify the parameters for the specific
scenarios (within each of the eight models) that will be explored
at the Tevatron using detailed Monte Carlo simulations.
The selected scenarios are particularly chosen to
sample the range of parameter space at the edge of Tevatron sensitivity.
In Sec. 5, we give details of the simulations and cuts that
we employ to analyze and isolate the different types of Tevatron discovery
signals. In Sec. 6, we give the numerical results for Tevatron
signals and backgrounds for the scenarios specified in Sec. 4.
By examining these results as a function of scenario location
in parameter space, we determine the portion of the parameter
space of each of our eight models for which a supersymmetric signal
will be detectable at the Tevatron. Substantial sensitivity
to specific boundary condition and unification scale choices emerges.
In Sec. 7, we survey the ability of LEP-II to explore the
parameter spaces of each of the eight models, and draw
comparisons to the Tevatron results.  The substantial complementarity
of the two machines is discussed. We present final remarks
and conclusions in Sec. 8.
Earlier analyses of selected Tevatron and/or LEP-II
signals for models similar to those considered here appear in
\REF\lnwz{
J. Lopez, D. Nanopoulos, X. Wang and A. Zichichi, \prdj{48} (1993) 2062.}
\REF\lnpwz{J. Lopez, D. Nanopoulos, H. Pois, X. Wang and
A. Zichichi, \prdj{48} (1993) 4062;
J. Lopez, D. Nanopoulos, G. Park, X. Wang and A. Zichichi,
CTP-TAMU-74/93.}
\REF\lnz{For a review, and further references, see
J. Lopez, D. Nanopoulos and A. Zichichi, CTP-TAMU-80/93.}
Refs.~[\lnwz-\lnz].

\vskip .15in
\noindent{\bf 2. String and Minimal Supergravity Models and Boundary
Conditions}
\vskip .075in

In string theory the unification scale and
the gauge couplings are determined dynamically at tree-level in terms of the
vacuum expectation values of the dilaton field, with
one-loop corrections coming from moduli field terms.
In general, gauge coupling unification
is not dependent on a grand unifying group, but instead takes the form:
$g_1^2k_1=g_2^2k_2=g_3^2k_3$,\Ref\Ginsparg{P. Ginsparg, \plbj{197},
139 (1987).}\ where the $k_i$ are
non-Abelian gauge factors called the Kac-Moody levels.
Phenomenologically consistent gauge coupling unification requires
$k_3=k_2={3\over 5}k_1$.  This is, in fact, the prediction
of the simplest and most attractive string theories, where one finds
Kac-Moody levels $k_3=k_2={3\over 5}k_1=1$.

The most obvious difficulty with the string approach is that
the string scale (at which the unification boundary conditions would naively
be expected to occur) is determined to be
$\mstring=0.7g_U\times 10^{18}\gev$,\Ref\stringscale{V. Kaplunovsky,
\npbj{307} (1988) 145.}\ \ie\ of order $M_P/\sqrt {8\pi}$
(where $M_P$ is the Planck scale --- $M_P=1.2\times
10^{19}\gev$ --- at which quantum effects of gravity must be considered),
as compared to the somewhat lower MSSM unification scale of
$\mgut \sim (2\times 10^{16})\gev$.
However, it is now known that there can be significant
threshold corrections associated with the infinite number of
Planck-scale states. Calculations in specific cases
\Ref\threshold{L. Dixon, V. Kaplunovsky and J. Louis,
\npbj{355}, 649 (1991); I. Antoniadis, K. Narain and T. Taylor,
\plbj{267}, 37 (1991); see also Ref.~[\stringscale].}\
show that these threshold corrections can effectively cause the couplings
at $\mstring$ to differ from one another in just such a manner as to yield
an effective unification at the lower $\mgut$ scale. It is also
possible that the unification point, $\mgut$, is {\it higher} than
`naively' predicted (\ie\ on the basis of the minimal
Standard Model (SM) particles plus associated superpartners plus Higgs
doublets) due to the presence of extra vector-like multiplets
with masses at an intermediate scale between $\mz$ and $\mgut$.
For appropriately chosen masses and representations these can modify the gauge
coupling running so as to yield coupling unification at $\mstring$.
This solution is apparently
{\it required} in the $4-D$ free-fermionic string formulation,
where the predicted threshold effects only
serve to increase the unification scale,\Ref\freethresh{S. Kalara,
J.L. Lopez and D.V. Nanopoulos, \plbj{269} (1991) 84.}\
and in certain orbifold constructions.
\Ref\orbicon{See, for example, Dixon \etal, Ref.~[\threshold].}\
For either resolution, the phenomenology of the string models
that yield the minimal supersymmetric model as the $N=1$ low-energy
supersymmetric sector is clearly worth examining. By comparing
the two approaches, we will gain a first indication of the sensitivity
of phenomenology to assumptions about unification-scale physics.

As noted earlier,
generically the MSSM has many independent parameters beyond those
required in specifying the Standard Model --- namely,
the magnitudes of soft-supersymmetry-breaking potential terms. The latter
are parameterized by scalar masses ($m$), soft Yukawa coupling
coefficients ($A$), gaugino masses ($M_a$, where $a$ denotes
the group), and the coefficient ($B$) specifying the
soft scalar Higgs field mixing term.
Fortunately, supergravity and superstring theory both provide insight
into the general structure of the soft terms. In supergravity,
rather mild assumptions yield universal soft-SUSY breaking, dramatically
reducing the number of parameters.
In string models there has been much progress in
classifying the possible SUSY-breaking sources; associated
`string-inspired' forms of soft-SUSY breaking emerge,
specified by a relatively small number of $\mgut$-scale parameters.
Rather universal soft-SUSY breaking can easily emerge.
Phenomenologically, FCNC constraints are most easily satisfied
if the soft-squark masses are, in fact, generation independent.
Of course, in principle a completely self-consistent,
calculable string theory should be able to predict {\it all} the
MSSM parameters. However, we are
far from realizing this goal, as numerous difficult and
unresolved issues in string theory still exist, including
the presumably non-perturbative effects that are
important in determining the true vacuum and the details of
supersymmetry breaking.

In this paper we shall consider two basic types of boundary conditions for
the soft-supersymmetry-breaking parameters at high energy scales.
For the first type, we adopt a
structure for the soft terms that emerges in a large number of
string models --- namely $M^0=-A^0=\sqrt 3 m^0$,
where $m^0$, $A^0$ and $M^0$ are the (common) values for the
soft scalar masses, $A$ coefficients, and $M_a^0$ gaugino masses, respectively.
(The superscript 0 denotes values at the high energy scale.)
These boundary conditions arise in the universal
dilaton-dominated limit of SUSY breaking in all string models,
but they are actually much more general, as we outline below.
As we shall see, the small value of $m^0$ compared with $M^0$ leads
to rather light spectra for the sleptons and sneutrinos of the
model. For comparison, we also consider the
extreme boundary conditions which yield the lightest possible slepton
sector: $m^0=A^0=0$. These latter boundary conditions,
supplemented by taking a universal value for the gaugino mass, $M_a^0=M^0$,
are commonly (but perhaps wrongfully) associated with the
``no-scale model'' label in the literature.

As well as specifying
the boundary conditions, the high energy scale at which they are applied
must be specified.  In this paper we study the consequences
of implementing the above two types of boundary conditions at
two different high energy scales: {\it i}) the standard
GUT scale $\mgut$ determined from gauge coupling unification
in the absence of any additional matter fields beyond those contained
in the minimal supersymmetric model; or, alternatively, {\it ii})
the substantially higher string scale.  In this latter case, extra
matter representations must be introduced --- our choices
will be detailed below.

\vskip .1in
\noindent{\it 2.1 String-motivated boundary conditions at $\mgut$}
\vskip .07in

\REF\kaplouis{V.S. Kaplunovsky and J. Louis, \plbj{306} (1993) 269.}
\REF\ibanez{A. Brignole, L.E. Ibanez, and C. Munoz, preprint FTUAM-26/93.}
A systematic analysis of soft-supersymmetry-breaking terms
in specific four-dimensional string theories has been presented in
Refs.~[\kaplouis,\ibanez]. In these references, the role of the dilaton ($D$)
and overall (\ie\ associated with the volume or size
of the manifold) moduli ($T$) fields is emphasized.
Starting with specific compactification
choices and the appropriate Kahler potential, explicit forms
for the soft-supersymmetry breaking terms can be derived when
the dilaton and/or moduli fields acquire non-zero vacuum expectation values
as a result of spontaneous supersymmetry breaking. In this approach,
the cosmological constant is {\it not} automatically zero; setting it
to zero further simplifies the soft-breaking boundary
conditions. Following the notation of Ref.~[\ibanez], the soft
parameters take particularly simple forms when expressed in
terms of the goldstino angle $\theta$, which specifies the extent to which
the source of supersymmetry breaking resides in the dilaton versus moduli
sector. If supersymmetry breaking
is dominated by the dilaton superfield then $\theta=\pi/2$, whereas
supersymmetry breaking dominance by the overall moduli superfield occurs in
the $\theta=0$ and $\pi$ limits.
For a given string model, the standard soft
parameters $m^0$, $M^0$, $A^0$, and $B^0$
can all be expressed in terms of just $\mgrtino$ (the gravitino mass)
and $\theta$.  For general values of $\theta$, the precise expressions
are model-dependent, although in the dilaton
limit of $\theta=\pi/2$ ($\sin\theta=1$), the soft terms take a
model-independent universal form (up to small corrections).

We consider a simplified subset of the models explored in
Ref.~[\ibanez]. First, we assume that the Kac-Moody
levels of the three gauge groups are related by
$k_3=k_2={3\over 5} k_1=1$. If we recall that $f_a=k_a S$ at tree level,
where $f_a$ are the inverse squared gauge coupling constants at the
string scale, we see that this choice will be consistent
with the experimentally observed coupling unification if
one neglects corrections arising from the difference between
$\mstring$ and $\mgut$.  Generally, one-loop threshold corrections
can alter this relationship somewhat (see Eq.~(2.12) of Ref.~[\ibanez]),
perhaps even allowing the apparent unification at $\mgut$ for
tree-level $k_i$ values (related as above)
to be consistent with threshold-corrected
unification at $\mstring$.
Second, we neglect any CP violating phases for the $A^0$ and $B^0$
parameters. Finally, in the case of orbifold compactification
we assume that all fields belong to the untwisted sector (\ie\
we take the modular weights to be $n_i=-1$
for all fields $i$; see Eq.~(3.9) of Ref.~[\ibanez]).
This set of choices is certainly the simplest possibility within
the context of the four-dimensional superstring models considered
in Refs.~[\kaplouis,\ibanez].

With these choices, the soft terms for both Calabi-Yau and
orbifold compactifications take the form:
$$\eqalign{m_i^{0\,2}=&\mgrtino^2\left[1-\kappa\cos^2\theta\right]\cr
           A^0=&-\sqrt 3 \mgrtino\sin\theta \cr
           M^0_a=&\sqrt 3\mgrtino\sin\theta +\mgrtino \cos\theta X_a\,, \cr
          }\eqn\compact$$
where $a$ labels the gauge group. In the absence of threshold corrections,
$\kappa=1$ and $X_a=0$. However, threshold corrections are generally present.
For orbifold compactifications
(with all $n_i=-1$) one-loop threshold corrections give rise to
$$\kappa=(1-{\delgs\over24\pi^2Y})^{-1},
\quad\quad X_a\propto b_a-k_a\delgs\eqn\oii$$
where $Y$ is computable in terms of the dilaton and moduli chiral superfields
and numerically is of order 4 or 5, $b_a$ is the one-loop $\beta$-function
coefficient for the particular group, and $\delgs$
is a model-dependent quantity (often a negative integer).
For Calabi-Yau compactifications, much less is known about the threshold
corrections, although it is quite probable that $\kappa$ is not precisely
1 nor $X_a$ exactly 0.\Ref\nilles{We thank H.P. Nilles for a discussion
on this point.}

Finally, we note that approximate results for the $B^0$ parameter were
obtained in Ref.~[\ibanez]. For example, if Higgs superfield mixing
appears only in the standard $\mu \hat H_1\hat H_2$ superpotential term
one finds
$$\bmu=\mgrtino(-1-\sqrt 3 \sin\theta -\cos\theta)\,.\eqn\bmudef$$
Another source of $B^0$ derives from an additional Higgs-mixing
term in the Kahler potential that can generally be present in Calabi-Yau
compactifications, but is not present for orbifold compactifications.
The resulting form of $B^0$ in the absence of the $\mu\hat H_1\hat H_2$
superpotential term is
$$\bz=2\mgrtino(1+\cos\theta)\,.\eqn\bzdef$$
If both sources of $B^0$ are simultaneously present, the resulting form
of $B^0$ as a function of $\theta$ would be more complicated.
And the above forms themselves were obtained only after a significant
number of approximations and assumptions. Thus, we shall leave $B^0$
as a free parameter subject only to the requirement that
the model be consistent with correct EWSB and phenomenological constraints.
However, we shall later describe the values taken on by $B^0$
within the allowed parameter space regions; we shall see that the models
would be very strongly constrained or eliminated altogether for
particular choices of $B^0$. In this regard, it is useful to note
that in the context of the above approximate forms,
$\bmu\leq 0$, while $\bz\geq 0$, with zero values only being reached
for $\theta=-\pi$.

{}From Eq.~\compact\ we extract the dilaton-dominated model
by setting $\sin\theta=1$, which gives identical $A^0$, $m_i^0$
and $M_a^0$ results for both Calabi-Yau and orbifold compactification
(as noted earlier) and is actually completely independent of the $n_i$ choices.
Indeed, to the extent that threshold corrections can be neglected
(in Ref.~[\ibanez] this is estimated to be true for $\sin\theta\gsim 0.05$),
the orbifold (with all $n_i=-1$)
and Calabi-Yau model results for $m_i^0$, $A^0$, and $M_a^0$ are
identical, and are indistinguishable from the strict dilaton-dominated model
after the rescaling $\mgrtino\rta \mgrtino\sin\theta$. If one could
trust one of the above quoted results for $B^0$, then it would provide some
discrimination between different $\theta$ values. However, the
uncertainty in $B^0$ makes it much more appropriate to allow
it to be a free parameter, in which case all models become equivalent
for the very large range of $\theta$ values such that
threshold corrections can be neglected. This large class of models,
specified by the boundary conditions
$$M^0=-A^0=\sqrt 3 m_i^0\,,\eqn\dilatonbc$$
will be denoted by the symbol $\dil$ (for dilaton-equivalent).
Leaving $B^0$ (as well as $M^0$) free yields (for a fixed
value of the top quark mass) a two dimensional parameter space
for the $\dil$ models, parameterized in terms of $\mgl$ and $\tanb$.
The $\dil$ phenomenology with $B^0$ left free will thus have a large
range of validity, and it is only in the $\sin\theta\sim 0$ case,
or by taking some of the $n_i$ different from $-1$ in the orbifold
models, that SUSY breaking can become model dependent in a manner that goes
beyond the boundary conditions that we shall employ.

\vskip .1in
\noindent{\it 2.2 Extreme minimal supergravity boundary conditions at $\mgut$}
\vskip .07in

The boundary conditions obtained
in the moduli-dominated limit (corresponding to $\sin\theta=0$ in
of Eq.~\compact) are quite different from the dilaton-equivalent
constraints of Eq.~\dilatonbc.
First, we note that the scale of $M_a$
is set by $X_a$, which therefore cannot be too small
on purely phenomenological grounds. More generally, the relative sizes
of the scalar masses and the gaugino masses are determined by the
relative magnitudes of $\kappa-1$ and $X_a$. Even in the specific
case of orbifold ($n_i=-1$) moduli-dominated models there are many
possibilities.  In the simplest case of $\delta_{GS}=0$,
$m_i^0=A^0=0$ and {\it non-universal}
values for the gaugino masses $M_a^0\propto b_a$ are required
at $\mgut$. In contrast, for significant
(negative, generally negative-integer) non-zero values
of $\delta_{GS}$, $m_i^0$ is typically substantially larger than $M_a^0$
in the moduli-dominated limit.  In the example of $\delta_{GS}=-5$
explored in Ref.~[\ibanez], gaugino masses become smaller than scalar masses
for $\sin\theta \lsim 0.05$. As noted there, this is not unlike the situation
obtained in explicit gaugino-condensation models.\Ref\decarlos{B. de Carlos,
J.A. Casas and C. Munoz, \plbj{306} (1993) 269.}\
The $\delta_{GS}=0$ case, although regarded as atypical (in that $S-T$
mixing will generally be present in the Kahler potential),
represents an interesting extreme case of boundary conditions
with specific non-universal gaugino masses at $\mgut$.

The extreme case of $\delta_{GS}=0$ with $\sin\theta=0$, yielding the
$M_a^0$ and $B^0$ as the only seeds of supersymmetry breaking, has
much in common with the so-called
`no-scale' scenario.\Ref\noscale{For a review and
references, see A.B. Lahanas and D.V. Nanopoulos, \prepj{145} (1987) 1.}\
The original motivations for no-scale models were i) to guarantee a vanishing
cosmological constant at the unification scale, and ii) to yield a
flat potential (in a scalar field direction) such that the electroweak
symmetry breaking (EWSB) scale
is to be generated dynamically. The no-scale scenario requires a Kahler
potential of a very specific form
(which is in fact realized in certain free-fermionic constructions
\Ref\lopyuan{J.L. Lopez, D.V. Nanopoulos and K. Yuan, CTP-TAMU-14/94.}).
Of course, soft-SUSY breaking
introduces a scale, but the underlying motivations and structure of
the `no-scale' model are least disturbed for a very simple and specific
set of soft-SUSY parameter boundary conditions; namely zero values
for the soft scalar masses and $A$ terms at $\mgut$, but non-zero values
for the gaugino masses.\refmark\noscale\ In the simplest models,
the chiral superfield density $f_{\alpha\beta}$ is proportional
to $\delta_{\alpha\beta}$, and gaugino masses take
on a universal value at $\mgut$.  The resulting boundary conditions are:
$$ m_i^0=A^0=0; \quad M^0_a=M^0\,.\eqn\noscalebc$$
As in the dilaton-equivalent case,
specific choices for the $B^0$ parameter are less motivated, and we will
allow any value for $B^0$ consistent with electroweak symmetry breaking
in the renormalization group approach.
We comment later on the values that $B^0$ actually takes on
in the parameter space regions that are consistent with other constraints;
$B^0=0$ is an interesting special case in that
gaugino masses then provide the sole seed for supersymmetry breaking.

The boundary conditions of Eq.~\noscalebc\ are best viewed
as specifying an extreme version of the
Minimal Supergravity model, for which we adopt the
the generic symbol $\ns$. As outlined above, they differ from
the boundary conditions that are obtained in the special moduli-dominated
$\delta_{GS}=0$ string theory limit where the $M_a^0$ are required
to be non-universal. To date, no detailed string model
has resulted in precisely the minimal-supergravity boundary conditions
of Eq.~\noscalebc\ with universal gaugino mass. Nonetheless, the $\ns$
boundary conditions have a long history and are the simplest that
can be devised which satisfy the basic constraints outlined below.
First, since setting $M^0_a=0$
would be inconsistent with experimental limits on gaugino masses,
Eq.~\noscalebc\ is the simplest choice for which
only one of $A^0$, $m^0_i$, $M^0_a$ is taken to be non-zero.
In addition, $m^0_i=0$ guarantees that the squark masses are sufficiently
degenerate after renormalization group evolution
to avoid flavor changing neutral current difficulties.
(Of course, a universal value for the $m^0_i$ as in the $\dil$ models,
also achieves the same end.)
Setting $A^0=0$ removes the CP violation that might
otherwise be present at a level inconsistent with
constraints on the neutron and electron electric dipole moments
should $A^0$ have a non-trivial phase with respect to $M^0$.
This is because the driving terms in the renormalization group equations
(RGE's) for the various $A$'s are proportional to the $M_a$'s
(with real coefficients) so that no phase for any $A/M_a$ can
be generated if $A^0=0$.
There could still be a non-trivial phase for $B^0/M^0$; in our analysis
we consider only real values for $B^0$.

\vskip .1in
\noindent{\it 2.3 Summary of $\mgut$ boundary conditions}
\vskip .07in

To summarize: in the dilaton-equivalent
models we require $M^0=-A^0=\sqrt 3 m^0$ at $\mgut$, while
in the minimal-supergravity models we require $m^0=A^0=0$ at the scale $\mgut$,
and a universal value, $M_a^0=M^0$ for the gaugino masses --- in both models
all values of $B^0$ are allowed that yield a correct pattern of
electroweak symmetry breaking and that are consistent
with existing experimental and phenomenological constraints
as detailed below.

So far, we have not discussed the $\mu$ parameter required to complete
the specification of the $\mgut$ boundary conditions.  Although the magnitude
of $\mu^0$ is determined in the RGE approach in terms of the other
$\mgut$-scale parameters by minimizing the scalar potential,
the sign of $\mu$ is undetermined; we will consider both the
positive and negative sign possibilities.  This leaves us with two $\dil$
and two $\ns$ models that we will explore phenomenologically.  We denote
these four models by $\dilp$, $\dilm$, $\nsp$, and $\nsm$.
We re-emphasize that by leaving the $B^0$ parameter free (subject
only to phenomenological and RGE constraints), the $\dil$ models
actually describe a broad class of superstring-motivated models in which
threshold corrections can be neglected, while the $\ns$ models are more
general than those in which $B^0=A^0-m^0$\refmark\noscale\ or $B^0=0$
is imposed.

\vskip .1in
\noindent{\it 2.4 Boundary conditions at $\mstring$}
\vskip .07in

As in the case of the gauge couplings, the
use of the boundary conditions \dilatonbc\ or \noscalebc\ at scale $\mgut$
as opposed to $\mstring$ can be questioned. Certainly, it is simplest
to presume that the soft-parameter boundary conditions apply at the same
scale at which the coupling constants begin their independent
evolutions from a common value.
A priori, it cannot be ruled out that the hidden-sector/threshold
corrections conspire so that both coupling constant equality
and universal values for the soft masses and the $A$ parameters
all apply at $\mgut$ rather than $\mstring$.
It is probably too early to even rule out the possibility
that the hidden-sector-determined scale at which
all evolution begins {\it is} significantly below $\mstring$.

However, as an alternative, one can presume that the threshold/hidden-sector
effects are small or (possibly) serve to push $\mgut$ even higher, in which
case it is most natural to presume that
that the coupling constants take a common value at
$\mstring$. To achieve coupling unification as this higher
scale, it is necessary
to introduce intermediate-to-GUT-scale (or gap) representations that
modify the gauge coupling running at high scales in precisely
the correct manner.\Ref\ibanezrun{L.E. Ibanez, \plbj{126} (1983) 196.}\
Of course, this approach implicitly assumes that
the nice meeting of the couplings in the MSSM (with minimal field content) is
quite accidental.  The minimal model along these lines is that suggested in
\REF\aekn{I. Antoniadis, J. Ellis, S. Kelley,
D.V. Nanopoulos, \plbj{272} (1991) 31; S. Kelley, J.L. Lopez and D.V.
Nanopoulos, \plbj{278} (1992) 140; S. Kelley, J.L. Lopez, D.V.
Nanopoulos, H. Pois, and K. Yuan, \npbj{398} (1993) 3.}
Ref.~[\aekn], in which vector-like heavy `quark' field representations
of the type
$$Q_L=(3,2,1/3), \quad Q_L^c=(\bar 3,2,-1/3),\quad D^c_R=(3,1,2/3),\quad
D_R=(\bar 3,1,-2/3),\eqn\gapfields$$
are introduced.
By inputing $\alpha_s$ and $\sin^2\theta_W$,
and requiring coupling unification at $\mstring$, the masses of these gap
fields can be determined. The masses required for unification are then
$m_{D_R}\simeq 10^7\gev$ and $m_{Q_L}\simeq 5\times 10^{12}\gev$.
Contributions to the $b_i$ are positive for both types of fields, thereby
increasing the slopes of the running $\alpha_i$ as a function of
the scale, which raises the unification scale as well as $\alpha_U$.
Indeed, these extra representations even have a natural home in the
flipped $SU(5)\times U(1)$ model:
the $Q_L$'s are the fermionic partners of the $X,Y$
gauge bosons and the $(3,2)$ components of the {\bf 10} Higgs representation;
the $D_L$'s are the partners of the Higgs triplet $(3,1)$ components
of the {\bf 5} and {\bf 10} Higgs representations.
(Of course, the $Q^c_L$ and $D^c_L$ are associated with the conjugate
states to those listed above.)
The flipped model mass matrices can
even yield masses for the $Q$ and $D$ fermions of
the required magnitude.\refmark\lnz\
We will discuss how the phenomenology changes upon employing
this approach and applying the dilaton-equivalent and
minimal-supergravity boundary conditions
of Eqs.~\dilatonbc\ and \noscalebc\ at $\mstring$.
The models so generated will be denoted by $\sdil^{\pm}$ and $\sns^{\pm}$
(for string-scale-unified dilaton-equivalent and string-scale-unified
minimal-supergravity), where, as before, the models are determined
by choosing just two parameters and the sign of $\mu$
(indicated by the superscripts).
The masses of the $D$ and $Q$ fields are not independent,
being entirely determined by the requirement that unification occur
at $\mstring$.

Of course, it must be admitted that the choice of unifying at $\mstring$
employing intermediate scale representations that are specific
to $SU(5)\times U(1)$ is somewhat arbitrary.
Indeed, as noted earlier, there is no need for the super-string to have
any group structure beyond the basic $SU(3)\times SU(2)_L\times U(1)$
of the SM. Nonetheless, we hope our
results will be representative of those that one would obtain for
other reasonably simple choices for the gap fields.

The main effect of raising the unification
scale from $\mgut$ to $\mstring$ is the increased slepton masses
(at a given gluino mass) due to the increased amount of evolution
lever arm.  This effect is clearly greater (on a percentage
basis) for the $\ns$ models where $m^0=0$ (and the entire slepton mass
at low energy derives from evolution) as compared
to the $\dil$ models for which $m^0$ is a significant fraction of $M^0$.
Important phenomenological differences occur if the mass hierarchy
of the sleptons, sneutrinos, charginos and neutralinos is altered.
We shall see that mass hierarchies are altered in going from
$\dilm$ to $\sdilm$ and from $\ns^{\pm}$ to $\sns^{\pm}$.
Some amusing patterns will emerge. In particular:
the $\dilm$ mass hierarchies are converted
to the ordering found for $\dilp,\sdilp$; the $\snsp$
hierarchies are quite different than for $\nsp$, and are
closely related to those for $\dilp$;
and the $\snsm$ hierarchies are very similar to those for
$\dilm$.  Thus, both the $\sdilp$ and $\sdilm$ will have phenomenology
similar to that found for $\dilp$, while $\snsm$ results
will be very like those for $\dilm$.  In combination, the
eight model scenarios investigated illustrate the vital role
played by precise unification boundary condition and scale choices
when supersymmetry breaking is dominated by gaugino masses.

\vskip .1in
\noindent{\it 2.5 Final introductory remarks}
\vskip .07in

Thus, the models we explore are motivated in two ways. On the theoretical
side, supergravity and string theory provide a strong motivation for the
types of boundary conditions considered. On the phenomenological side,
the boundary conditions employed
are distinguished by the rather low masses obtained for the
sleptons.  This latter provides a very interesting alternative to
models in which boundary
conditions with large $m^0$ arise, such as gaugino-condensate models and
models based on the above-mentioned moduli-dominated string scenarios
(with $\delta_{GS}\neq0$). Light sleptons will turn out to have
many crucial phenomenological consequences (such as two-body decays
of charginos and neutralinos to slepton-lepton final states)
that dramatically alter the phenomenology as compared to models where
sleptons have mass of order the gluino mass or higher
(so that charginos and neutralinos tend to decay via three-body
modes to virtual or real $W,Z$ plus lighter gaugino).
We shall see that the predicted slepton masses are generally
sufficiently smaller than the gluino mass that slepton-pair production
at LEP-II can easily be a much deeper probe of the model parameter
space than the Tevatron, whereas the reverse can easily be true
if sleptons are heavy.  Light sleptons also result in many special
situations and much greater variability in the discovery potential
at the Tevatron. Signals for supersymmetry will often be more likely to
arise from neutralino/chargino-pair and slepton-pair production
than from gluino-pair, gluino-squark, or squark-pair production.
(Squarks generally turn out have fairly large mass, except
possibly the lighter stop squark, as a result of the
$\alpha_s$ terms in the RGE's, which are absent for the sleptons.)
Experimental searches at the Tevatron should pay more attention
to the types of discovery modes that we shall delineate in the following
sections.

\vskip .15in
\noindent{\bf 3. Model Parameter Space Constraints}
\vskip .075in

The eight different models that we shall explore were delineated
in the Introduction.  They will be denoted by
$\dilp$ (dilaton-equivalent, $\mu>0$), $\dilm$ (dilaton-equivalent, $\mu<0$),
$\nsp$ (minimal-supergravity, $\mu>0$), $\nsm$ (minimal-supergravity, $\mu<0$),
$\sdilp$ (string-scale dilaton-equivalent, $\mu>0$),
$\sdilm$ (string-scale dilaton-equivalent, $\mu<0$),
$\snsp$ (string-scale minimal-supergravity, $\mu>0$)
and $\snsm$ (string-scale minimal-supergravity, $\mu<0$).
As noted in the introduction, all eight of these models
are entirely determined in terms of just two parameters (in addition to
the sign of $\mu$), given a definite value of the
top quark mass (which fixes the Yukawa couplings of the theory
when combined with our other boundary conditions).
We will employ $\mt(\mt)=170\gev$, corresponding
the $\mt(pole)=178.3\gev$ for all our computations, as being
representative of the best fit value for LEP data
\Ref\blondel{See, for example, A. Blondel, presented at
the Teupitz meeting on ``Physics at LEP-200 and Beyond''.}
and a not-unlikely value for explaining the excess events at
the Tevatron.\Ref\cdftop{CDF Collaboration, FERMILAB-PUB-94/097-E.}
We note that we do not require unification of the bottom
and tau Yukawa couplings at $\mgut$.  Typically, their ratio
at $\mgut$, $\rbtau$, is within 15\% of unity in the models
being considered. Although it would greatly simplify
our considerations by reducing
the parameter space to just a single dimension, requiring
absolutely precise Yukawa unification may well be an
artificially strong constraint.

Various choices for the two model-determining parameters
can be considered.  We have found it convenient to employ
$\mgl$ and $\tanb$, where $\mgl\equiv\mgl(pole)$
is the mass of the physical gluino state, and $\tanb$ is the ratio of the
Higgs doublet field vacuum expectation values.
Of course, in employing these two low-energy parameters
we are using a bottom-up approach to the renormalization group equations,
\REF\pokorski{M. Olechowski and S. Pokorski,
\npbj{404} (1993) 590.}
as pioneered in Ref.~[\pokorski].  The RGE's are solved by
employing one-loop evolution equations, adopting $\alpha_s(\mz)=0.12$,
$\alpha_{QED}=1/127.9$, $\mb(\mb)=4.25\gev$ and $\mt(\mt)=170\gev$.
Evolution is performed between $\mgut$ or $M_S$ and $\mz$.
The resulting values of $\mgut$ and $\alpha_U$
are $\mgut=2.39\times 10^{16}\gev$, $\alpha_U=0.0413$ for the $\dil$, $\ns$,
models and $\mgut=1\times 10^{18}\gev$, $\alpha_U=0.0551$ for
the $\sdil$ and $\sns$ models.
At one-loop, this procedure predicts $\sin^2\theta_W= 0.2305$,
\ie\ outside the errors on the experimentally
preferred value of 0.2324;\refmark\blondel\ however,
it is well-known\Ref\twoloopalphas{P. Langacker and N. Polonsky,
\prdj{47} (1993) 4028.}\
that the two-loop corrections to the RGE's
bring $\sin^2\theta_W$ into much closer agreement
($\sin^2\theta_W\simeq .2335$)
with the experimental result for the chosen value of $\alpha_s(\mz)=0.12$.
At one-loop, we prefer to employ this value for $\alpha_s(\mz)$, that is
more in the center of its experimentally allowed range,\refmark\blondel\
as opposed to the lower value of $\alpha_s(\mz)=.11$ that at one-loop would
yield $\sin^2\theta_W=0.2324$, due to its influence on the RGE equations
for the strongly interacting sparticle masses and Yukawa couplings.
In any case, uncertainties associated with $\mgut$ or $\mstring$ boundary
conditions and threshold corrections are surely larger than those associated
with these approximations. Finally, we note that
in evolving the `$\lambda$' parameters entering
the Higgs potential,\Ref\hablambda{H.E. Haber and R. Hempfling,
\prdj{48} (1993) 4280.}\  the various
sparticles are decoupled at their respective mass scales; this
is accomplished by an iterative method.

The first step in our analysis
is to determine the allowed region of the \mgltb\ parameter
space for each of the models.
The constraints that we shall apply are the following, not all
of which turn out to be important.
\pointbegin
We require a neutral LSP. This requirement determines
the upper limit on the mass of the gluino
at fixed $\tanb$ that arises in the $\nsp$ and $\nsm$ models; if $\mgl$ becomes
too large, $\mcnone$ exceeds $\mstauone$.
\point
We require that the $\hl$ and $\ha$ of the model not be visible at LEP.
This is not constraining, due to the large value of $\mt(\mt)=170\gev$
that we employ.
\point
We require that all sleptons be heavier than $\mz/2$, since
sleptons are not observed in $Z$ decays.
\Ref\leplim{
D.~Decamp {\it et.al.} (ALEPH Collaboration), \plbj{236} (1990) 86;
P.~Abreu {\it et.al.} (DELPHI Collaboration), \plbj{247} (1990) 157;
O.~Adriani {\it et.al.} (L3 Collaboration), CERN-PPE-93-31 (1993);
M.~Akrawy {\it et.al.} (OPAL Collaboration), \plbj{240} (1990) 261;
for a review, see G. Giacomelli and P. Giacomelli, CERN-PPE/93-107 (1993).}\
Requiring $\msnu>\mz/2$ determines the lower $\mgl$ boundary
for the $\snsm$ and $\dilm$ models at all $\tanb$ values, and
for the $\nsp$, $\snsp$ and $\sdilm$ models for $\tanb\lsim 2.3$, 4.5,
and 10.5, respectively.
It also determines the upper bound on $\tanb$ in the $\dil$, $\sdil$
and $\sns$ models --- at large $\tanb$
splitting between the $\stau$'s becomes sufficiently large that
the $\stauone$ is pushed to a mass below $\mz/2$.
\point
We require that the lightest chargino, the $\cpone$, be heavier than
$\mz/2$, since chargino pairs are not observed in $Z$ decays.\refmark{\leplim}\
This requirement determines the lower $\mgl$ bound for the $\dilp$ and $\sdilp$
models, and for the $\nsp$, $\snsp$ and $\sdilm$ models for $\tanb\gsim 2.3$,
4.5, and 10.5, respectively.
\point
We require that $\mstopi>\mz/2$, where $\stopi$ is the lighter
of the (rather widely split) stop squark mass eigenstates.\refmark\leplim\
This requirement is not constraining in the cases studied.
\point
\REF\cdfsusy{F. Abe {\it et. al.}, \prlj{69} (1992) 3439.}
\REF\dzsusy{See the talk by M. Paterno, SUSY'94 Workshop, Ann Arbor, MI, 1994.}
We require $\msq>100\gev$, for all squarks other than the $\stop$.
This is only a rough lower bound from CDF/D0 data\refmark{\cdfsusy,\dzsusy}\
but is, in any case, not constraining for the models we study.
\point
Similarly, we note that a CDF/D0-like requirement of $\mgl>120\gev$ is not
constraining. In fact, the slepton, chargino, and LSP boundary conditions
require that the minimum value of $\mgl$ is always somewhat above $200\gev$
in the models considered.
\point
We require that the net contribution from new states to
the $Z$ width be smaller than 0.028 GeV, and that any additional
contribution to the $Z$'s invisible decay width be
$<0.018\gev$.\refmark\leplim\
Neither requirement is constraining for the models considered.
\point
We also demand that the EWSB potential minimum
for any acceptable solution to the RGE equations be a true global
minimum and that the potential be bounded at $\mgut$ or $M_S$.
\point
We require that the top quark Yukawa coupling remain perturbative
as defined by $h_t\leq 3$.  For Yukawa coupling larger than this
the two-loop corrections to the one-loop renormalization group
equations become large and the perturbative approach begins to break down.
This requirement determines the lower boundary, \ie\
smallest allowed value of $\tanb$, for all the models.

\noindent
A possible further constraint on the models, that we shall not directly
implement, derives from the fact that the $\cnone$ can provide
\Ref\darkmatter{See M. Drees and M. Nojiri, \prdj{47} (1993) 376;
S. Kelley, J.L. Lopez, D.V. Nanopoulos, H. Pois and K. Yuan,
\prdj{47} (1993) 2461;
L. Roszkowski and R. Roberts, \plbj{309} (1993) 329;
R. Arnowitt and P. Nath, \plbj{299} (1993) 58 and ERRATUM \plbj{307}
(1993) 403; G. Kane, C. Kolda, L. Roszkowski and J. Wells, UM-TH-93-24
and references therein.}\
a significant dark-matter density in the early universe.
As is well-known,\refmark\darkmatter\
if the $\cnone$ becomes too heavy, and if the cross section for
the annihilation of $\cnone$ pairs is not large, then the universe
can be overclosed or `too young'.
However, in the models we consider the sleptons
and sneutrinos are generally rather light, which tends to enhance
the annihilation cross sections.  Thus, we expect much of the
parameter space illustrated to remain allowed by even a rather
stringent constraint on dark matter relic density.
This is illustrated for example in the investigations
of Ref.~[\lnpwz]. At most, large $\mgl$ values, \ie\
beyond those relevant for Tevatron searches, would be eliminated
by imposing this constraint.

We also do not implement proton decay constraints. While full
gauge group unification at $\mgut$ can lead to difficulties with proton decay,
\foot{For example, in $SU(5)$ unification proton
decay is frequently too rapid.\Ref\rapidref{S. Hisano, H. Murayama
and T. Yanagida, \npbj{402} (1993) 46; J.L. Lopez, D.V. Nanopoulos,
and H. Pois, \prdj{47} (1993) 2468; R. Arnowitt and P. Nath,
\prlj{69} (1992) 725; P. Nath and R. Arnowitt, \plbj{287} (1992) 89.}}
such full unification is not typical of
string theories. In string theories, we have noted earlier
that coupling constant equality at $\mgut$ is
instead a result of the simplest and most attractive choices
for the Kac-Moody levels, $k_i$. Indeed, many string models
with only the minimal SM group structure have been constructed.
\Ref\smstring{See, for example, L.E. Ibanez, D. Lust, and G.G. Ross,
\plbj{272} (1991) 251; A. Faraggi, D.V. Nanopoulos, and K. Yuan,
\npbj{335} (1990) 347; A. Faraggi, \plbj{278} (1992) 131.}\
In the absence of full gauge-group
unification, the $X$ and $Y$ gauge bosons and the especially
troublesome Higgs triplets need not be present with the result
that there is no definitive constraint on the models coming from proton decay.
Of course, the $\sdil$ and $\sns$ scenarios
fit nicely into the flipped $SU(5)\times U(1)$ model, for which
there need not be a problem with proton decay (because of the large
scale at which unification takes place)
despite the fact that true gauge-group unification occurs.

\FIG\boundaries{We plot the boundaries of allowed \mgltb\
parameter space for the dilaton-equivalent, minimal-supergravity, and
string-scale-unified dilaton-equivalent and minimal-supergravity scenarios
for both $\mu>0$ and $\mu<0$.  Values of $\rbtau$ for a given
$\tanb$ are given on the right-hand axis. The numbers indicate
the different cases considered for each scenario (see text). They are
positioned so that the actual \mgltb\ value associated with a point
is at the lower left-hand side of the number.}
The allowed regions of \mgltb\ parameter
space for the eight models obtained by imposing
these constraints  are displayed in Fig.~\boundaries.
Note that these constraints alone do not serve to determine a right-hand
boundary in the $\dil$, $\sdil$ or $\sns$ scenarios.  Presumably, it would be
unreasonable to consider solutions with $\mgl>1\tev$ purely
on the aesthetic ground that such a large gluino mass would
bring into question the original naturalness motivation for the MSSM.
We have confined ourselves to $\mgl<800\gev$ simply because
of our focus on the Tevatron in this paper. As noted earlier,
we have not imposed unification of the $b$ and $\tau$ Yukawa couplings
at $\mgut$.  However, a choice for $\tanb$
(along with $\mt(\mt)=170\gev$ and $m_b(m_b) =4.25$)
determines the ratio, $\rbtau(\mgut)$, independently of $\mgl$.
This ratio is given for each scenario along the right hand axes
in Fig.~\boundaries.

It is also of interest to outline the values taken on by $B^0$
in the allowed regions of Fig.~\boundaries\ for each of the eight models.
For the $\dilp$ model, $B^0<0$ throughout the allowed region
of Fig.~\boundaries a. Adopting the approximations
of Eqs.~\bmudef\ and \bzdef, this would exclude a Calabi-Yau model with
only $\bz$. However, $B^0/\mgrtino=-2$ and $-(1+\sqrt 3)$
both fall within the allowed parameter space (at $\tanb\sim 4$ and $2.7$,
respectively --- note that $B^0$ becomes less negative as $\tanb$ increases).
This means that a $\bmu$ source would be entirely consistent for all but
$\theta$ very near 0 (for which our boundary conditions are not appropriate
in any case). The results for the $\sdilp$ model are essentially identical.
For the $\dilm$ and $\sdilm$ models, $B^0>0$ everywhere in the allowed
parameter space depicted in Fig.~\boundaries a and b.  This requires
the presence of a $\bz$ mixing source; in other words, orbifold
compactifications are excluded in the context of the approximations
of Eqs.~\bmudef\ and \bzdef. In the $\dilm$ and $\sdilm$ models $B^0$ increases
as $\tanb$ decreases; however, the $\theta\sim0$ limit of
$\bz/\mgrtino\sim 2$ falls at $\tanb$
values below 2, which are excluded by non-perturbative top-quark
Yukawa coupling behavior for $\mt=170\gev$. (For lower $\mt$
values this limit {\it is} reached within the allowed domain.)
Turning to the $\ns$ and $\sns$ models, the basic features of $B^0$
are easily summarized. For $\nsp$ and $\snsp$, $B^0<0$ throughout
the allowed parameter spaces shown in Figs.~\boundaries c and d.
$B^0$ becomes less negative as $\tanb$ increases, but never reaches 0,
\ie\ the value consistent with $B^0=A^0-m^0$ for our choice of $A^0=m^0=0$.
For the $\nsm$ or $\snsm$ models, $B^0$ goes from positive to negative values
as $\tanb$ increases, passing through zero
at $\tanb\sim 7-9$ or $\tanb\sim 5-9$, respectively (larger $\tanb$
for larger $\mgl$). Thus, the strict no-scale boundary conditions
of $m^0=A^0=B^0=0$ are only possible for $\mu<0$.
As stated earlier, we shall not place
any restriction on $B^0$ in our phenomenological analyses.
The above $B^0$ results are presently only of passing
theoretical interest, but could become useful should predictions for $B^0$
become more certain at some future date.

The phenomenology of the different models is largely determined
by the masses of the super particles. Thus,
before turning to the specific scenarios that we shall examine
with regard to detection at the Tevatron or at LEP-II, it is useful to
illustrate the basic structure for the eigenstate masses that
emerge from the four types of models being considered.
\FIG\masses{We plot the masses of the various superpartner particles
in terms of the ratio $m_i/\mgl$ ($i$ denoting a particular sparticle)
as a function of $\mgl$.  The scatter in the points indicates the
variation as $\tanb$ is allowed to vary at fixed $\mgl$.
The band corresponding to a given particle type $i$ is labelled,
according to its relative position at {\it large} $\mgl$, on the r.h.s.
of the graph.}
Since, for a given model, the only mass scale is $\mgl$,
it is not surprising that all masses when plotted in ratio to $\mgl$
exhibit approximate scaling.  Only the variation with $\tanb$
(through the limited range allowed by the parameter space boundaries)
yields any scatter.  These scaling laws are illustrated in Fig.~\masses,
where the $m_i/\mgl$ values exhibit a well-defined band for
a given choice of sparticle type, $i$.
Note that at large $\mgl$ the $m_i/\mgl$ ratio for a given sparticle
is essentially independent of $\tanb$.  The gaugino masses
exhibit the standard relations\Ref\ghino{J.F. Gunion
and H.E. Haber, \prdj{37} (1988) 2515.}\
$\mcnone\sim M^\prime$ and $\mcpmone\sim M$ that arise whenever
$|\mu|>M^\prime,M$, where $M^\prime$ and $M$ are the $U(1)$ and $SU(2)$
low-energy soft masses respectively. ($\mcntwo$ is not plotted;
as expected in the above limit, $\mcntwo$ is always very close
to $\mcpmone$.) These mass limits
are correlated with the fact that the $\cnone$ and $\cntwo$
become primarily bino and wino in the large $|\mu|$ limit,
a fact which we shall see has some rather important phenomenological
consequences.

The crucial role of the mass hierarchies illustrated
in Fig.~\masses\ is in determining the production rates
and decay chains that are the dominant ingredients in the
phenomenological consequences of a particular choice for the model,
and the values of $\tanb$ and $\mgl$ within the given model.
There are important similarities as well as important differences
between the models in this respect. As noted in the introduction,
a very important point to note is that since the soft scalar masses $m_i$
are either zero at $\mgut$ (for the minimal-supergravity scenarios)
or at least smaller than the common gaugino
mass by a factor of $\sqrt 3$ (for the dilaton-like scenarios),
the scalar partners of the SM fermions acquire mass
largely as the result of evolution from the unification scale $\mgut$
down to $\sim\mz$. Since the evolution of the sleptons
is much slower than that of the squarks (there being no strong
interaction terms driving them away from zero mass), sleptons
will always be very much lighter than squarks in these scenarios.
Even the squarks only reach masses as large as the gluino mass
for the higher unification scale at $M_S$.
In the case of the GUT-scale-unified minimal-supergravity models, the sleptons
are quite light, and indeed not much heavier than the $\cnone$.
This is why limits on the slepton mass set the upper and lower
boundary on $\mgl$ in the $\ns$ models. In the $\sns$ models,
the sleptons and squarks move up in mass as a result of the
increased amount of evolution arising from the larger difference
between $\mstring$ and $\mz$. For the $\dil$ models, slepton
masses are again larger than in the $\ns$ models, but now as a result
of the non-zero seed scalar masses ($m_i\neq 0$) at $\mgut$.
Indeed, the slepton masses are not very different in the $\dil^{\pm}$
and $\sns^{\pm}$ scenarios.

However, there are subtleties in comparing slepton to ino
masses that have considerable phenomenological importance.
For $\mgl\lsim 500\gev$, the region of interest in our Tevatron study,
the mass hierarchies for $\dilp$ and $\snsp$ are such that
$\cntwo\rta \nu\snu$ and $\cpone\rta l\snu$
decays are generally forbidden,
whereas for $\dilm$ and $\snsm$ these decays are generally allowed.
We shall see that this results
in significant similarities between the $\sns$ and $\dil$ phenomenologies
for both signs of $\mu$. The $\sdilp$ and $\sdilm$ hierarchies are
such that the lightest chargino is generally lighter
than the lightest slepton, the $\snu$ (except at the lowest allowed $\mgl$
values).  As a result $\cntwo\rta \nu\snu$ (recall that
$\mcntwo\simeq\mcpmone$) and $\cpone\rta l\snu$
are again forbidden (except for quite low $\mgl$ and $\tanb\lsim 10.5$
in the $\sdilm$ case).
This will imply some similarity of $\sdilp$ and $\sdilm$
phenomenology to that of the $\dilp$ model.

In contrast, since the squark
masses in all the models considered derive mainly from evolution,
the squark masses in the $\dil$ and $\ns$ models are of similar size
and somewhat smaller than those predicted in the $\sdil$ and $\sns$ cases.

In summary,
we see that the gluino mass is the largest mass of all sparticles in these
scenarios, and hence may not be particularly relevant for phenomenology,
contrary to typical expectations. Squark masses are typically $\sim10-50$ GeV
below gluino masses. Sleptons are amongst the lightest SUSY particles;
hence the charginos and neutralinos often decay
via two-body modes to {\it e.g.} slepton-lepton, instead of the usually
expected three-body decays. We shall see that this can have a substantial
influence on the types of collider signatures expected.
In addition, we will find that the sleptons are generally sufficiently
light that a significant portion of parameter space can be probed
via slepton-pair production at LEP-II.

\vskip .15in
\noindent{\bf 4. Scenarios}
\vskip .075in

The allowed regions in \mgltb\ parameter space for the
minimal-supergravity and dilaton-equivalent scenarios (for both
$\mu>0$ and $\mu<0$) are illustrated in Fig.~\boundaries.
In order to explore the ability of the Tevatron to probe these eight
distinct scenarios, we have selected a series of points
in each of the eight allowed parameter spaces that comprise a representative
sample of cases that might also have some
chance of being accessible to the Tevatron.
The sampled points are numerically labelled in Fig.~\boundaries, and will be
referenced by $\dilp_i$, $\dilm_i$, $\sdilp_i$, $\sdilm_i$,
$\nsp_i$, $\nsm_i$, $\snsp_i$, and $\snsm_i$, where $i$ is the
numerical index indicating the sampled point within a given scenario.
The locations of all these scenarios in the eight parameter spaces are
shown in Fig.~\boundaries\ using the numerical label for a given scenario.
In the $\sdil$ and $\sns$ scenarios, we have focused primarily
on points that lie near the limit of experimental sensitivity at the Tevatron.

The phenomenological consequences of a given point within one of the
eight scenarios is largely determined by the masses and decay
branching ratios of the superpartner sparticles.
\TABLE\massest{}
In Tables~\massest a--d (for $\dil$, $\sdil$, $\ns$,
$\sns$ models, respectively)
we give the scenario label and $\mgl$, $\tanb$ values
for a given scenario point, along with the masses of the
lightest two neutralinos, $\cnone,\cntwo$, the lightest chargino, $\cpone$,
the left-handed slepton, $\slepl$, the right-handed slepton, $\slepr$,
the sneutrino, $\snu$, the first and second family squarks, denoted
generically by $\sq$, and the lighter of the two stop eigenstates
as obtained after diagonalization, $\stopi$. The Higgs masses
$\mhl$ and $\mha$ are also tabulated. Regarding the $\hl$,
our procedure is to compute $\mhl$ using the one-loop effective
potential corrections (including stop and sbottom mass splitting effects)
\REF\oneloophl{J. Ellis, G. Ridolfi and F. Zwirner, \plbj{257} (1991)
83 and \plbj{262} (1991) 477;
A. Brignole, J. Ellis, G. Ridolfi and F. Zwirner, \plbj{271}
(1991) 123; for explicit formulae, see M. Bisset, Ph. D. thesis,
U. of Hawaii, (1994).}
in the manner of Ref.~[\oneloophl] after having carried out the RGE
evolution in the manner described earlier. The $\hl$ masses obtained
in this way are somewhat higher (slightly lower) in the $\mu>0$ ($\mu<0$)
cases than those emerging directly from the RGE's.

\TABLE\brs{}
Some important branching ratios for each scenario point
appear in Tables~\brs a--d. In the $\dil$, $\sdil$ and $\sns$ scenarios the
two-body decays for $\cpone$ and $\cntwo$ into $\slepl$ are not allowed
or are completely negligible, so we do not show these.
Also, in the minimal-supergravity scenarios, the $\snu$ always decays
invisibly to $\cnone\nu$; thus, branching ratios for the $\snu$
are not tabulated for the $\ns$ case.
In the $\dilp$ and $\sdil^{\pm}$ scenarios,
and in one case each for
$\dilm$ and $\snsp$, there are significant visible decays
of the $\snu$, as indicated.
Finally, the $\dilmiv$
case is rather special in that the two-body decays
$\cpone\rta\cnone \wp$ and $\cntwo\rta\cnone \hl$
(where $\hl$ is the light CP-even Higgs boson) are allowed and
dominant (though not listed).

The most crucial distinction
amongst models however, derives from the fact that $\cpone$ and $\cntwo$ decays
tend to be saturated by the modes $\cpone\rta \snu l$ and $\cntwo\rta \snu \nu$
when kinematically allowed; this is quite characteristic of
the $\nsp$, $\nsm$, $\dilm$ and $\snsm$ scenarios.  The dominance
of the $\snu\nu$ channel in the case of the $\cntwo$
occurs despite the fact that the $l\slepr$ channel generally has
at least as much kinematic phase space. This is because the $l\slepr$
couples only to the bino component of the $\cntwo$, which is quite
small whenever $|\mu|>M$, as noted earlier.  Indeed, it should be noted that
the $\cntwo$ approaches a pure wino state much more rapidly for
large values of $\mu<0$ than for $\mu>0$.  Thus, the $l\slepr$
decays of the $\cntwo$ are particularly suppressed for the $\mu<0$ cases.
The $\snu\nu$ channel dominance is important phenomenologically since the
$\snu$
decays entirely to the invisible $\cnone\nu$ channel
whenever $\cpone\rta \snu l$ and $\cntwo\rta \snu \nu$
decays are allowed.
The result is a depletion of the visible event rate from $\cntwo$
decays, especially for the $\mu<0$ scenarios where the
$l\slepr$ branching ratio is particularly suppressed.

The net effect of these branching ratios on signatures is difficult to
deduce without a complete simulation. For instance, in the search for
trilepton events from $p\bar p\rta\cpmone\cntwo\rta 3\ell+\etmiss$,
the chargino and neutralino branching fractions can vary considerably, leading
to wide ranges of signal rates. In addition, kinematic effects can be
important.
For instance, if $\cntwo\rta\slep +\ell$, the final state $\ell$ may be too
soft to pass detector requirements, even if the branching ratio is large.

We shall see that the $\dilm,\snsm$ scenarios
are distinctly more difficult to detect than the $\dilp,\snsp$ ones.
However, the $\nsm$ scenarios will turn out to be as easily probed as the
$\nsp$
scenarios.  This is because of the `inverted' mass hierarchy,
$\mcpmone>\mslepl$ for large $\mgl$ in $\ns$ models compared
to $\dil$ and $\sns$ models. The inversion allows for $\cntwo\rta\slepl l$
decays (for which the coupling does not go to zero as the $\cntwo$ approaches
a pure wino state, although it is small).
The resulting production and decay chain
$\cpmone\cntwo\rta \slepl\nu+\slepl \ell$, followed by
two $\slepl\rta \ell\cnone$
decays, yields $3\ell$ states at a significant rate.
Meanwhile, the similarity between the $\sdil^{\pm}$ and the $\dilp$
mass hierarchies and decays implies that the $\sdil^{\pm}$ models
will be more or less as easily probed as the $\dilp$ case.

Before closing this section, we note that some of the above
discussion is peculiar to the $\mgl$ mass range relevant to Tevatron
exploration. In particular, for high enough $\mgl$ values the $\snu$ becomes
heavier than the $\cpmone$ for the $\dilm$ and $\snsm$ cases,
and thus would decay visibly, exactly as for
the $\dilp$ and $\sdil^{\pm}$ model cases explored here.
Correspondingly, the $\cntwo\rta \snu \nu$ and $\cpone\rta l\snu$ two-body
decays become kinematically disallowed. Three-body
decay channels would play a more prominent role, as in the
$\dilp$ and $\sdil^{\pm}$ cases. The final leptons would generally be harder
as a result (although perhaps less numerous). Overall, there could be a
temporary increase in the tri-lepton rate after cuts (as $\mgl$ is increased).
(This is, in fact, the source of the greater observability that
we shall find for the $\dilp$ and $\sdil^{\pm}$
as compared to the $\dilm$ and $\snsm$ models.)
This illustrates how phenomenological considerations
could well change significantly in moving to either higher luminosity
or higher energy at the Tevatron.  And certainly discovery potential
at the LHC cannot be extrapolated from the results we shall present
here.\Ref\furtherwork{Assessment of these situations is in progress.}
In general, the phenomenological complexity of the types of models
considered is substantial because of the delicate
cross-over's in masses and decay modes.

\smallskip
\noindent{\bf 5. Simulation and Selection Cuts}
\smallskip

 \TABLE\prods{}
 \TABLE\xsecsbkgnd{}
 \TABLE\xsecs{}

To simulate signal and background events at the Tevatron collider, we
use the event generator program
ISAJET 7.07.\Ref\isaref{F. Paige and S. Protopopescu, in {\it Supercollider
Physics}, p. 41, ed  D. Soper (World Scientific, 1986); H. Baer, F. Paige,
S. Protopopescu and X. Tata, in {\it Proceedings of the Workshop on Physics
at Current Accelerators and Supercolliders}, ed. J. Hewett, A. White and
D. Zeppenfeld, (Argonne National Laboratory, 1993).}
ISAJET 7.07 has been set up to perform a reasonable simulation of
supergravity models, provided $\tanb\lsim 10$. For larger $\tanb$ values,
the approximate degeneracy of sleptons is badly broken, and there are
large mixing effects for $\sbot$ and $\stau$ states. For two
simulations at large $\tanb$ --- the $\dilpvi$ and $\dilmvi$ cases ---
two-body decays of charginos and neutralino to real staus are dominant: in
these cases we force the relevant decays to occur with 100\% branching.

Briefly, for a given set of weak scale MSSM parameters, ISAJET calculates
branching fractions for all sparticle decay modes. ISAJET then generates
all SUSY particle production processes according to their relative
cross sections, and decays the various SUSY and SM particles via the
calculated or measured branching ratios. Initial and final state QCD
radiation is included, as is hadronization of quarks and gluons. Underlying
event soft-scattering is included as well.

The relative production cross sections for a variety of superparticle-pair
channels are given in Tables~\prods a--d  for all of the numbered scenario
cases
appearing in Fig.~\boundaries\ and in Tables~\massest\ and \brs.

We see immediately that $\gl\,\gl$, $\gl\,\sq$ and $\sq\,\sq$ production
processes often have
low rates compared to other SUSY-pair production processes,
especially for large values of $\mgl$.
Instead, the dominant production
processes are the cumulative $\chitil\chitil$ subprocesses, especially
$\cpmone\cntwo$ and $\cpmone\cmpone$.
Furthermore, the light slepton and sneutrino masses
characteristic of these models results in a substantial rate for
$\slep\,\snu+\snu\,\snu$ and $\slep\,\slep$ production.
Finally, there is significant rate
for the associated production final states $\chitil\,\gl$ and $\chitil\,\sq$,
which comprises the remainder of the event rate.

To model collider detector effects, we employ the toy calorimeter
simulation ISAPLT. We assume calorimetry extends over the central region
out to rapidity of $|\eta |<4$, with cell size $\Delta\eta\times\Delta\phi =
0.1\times 0.1$. We take hadronic energy resolution to be $70\%/\sqrt{E_T}$,
and electromagnetic resolution to be $15\%/\sqrt{E_T}$. Jets are coalesced
within cones of $R=\sqrt{\Delta\eta^2 +\Delta\phi^2}=0.7$, using the ISAJET
routine GETJET; clusters with $p_T>15$ GeV are labelled as jets.
Muons and electrons are classified as isolated if they have $p_T(\ell )>8$ GeV,
$|\eta (\ell )|<3$,
and there is less than $p_T(\ell )/4$ GeV in a cone of $R=0.4$
about the lepton direction. For all supersymmetric event topologies, we
require at least $\etmiss >20$ GeV.

\REF\bkt{H. Baer, X. Tata and J. Woodside, \prdj{41} (1990) 906;
H. Baer, C. Kao and X. Tata, \prdj{48} (1993) 2978.}
Multi-lepton signals for gluinos and squarks have been examined in
Ref.~[\bkt]. We follow approximately the cuts given there.
The cross sections examined include the following.
\pointbegin Multi-jet plus missing energy ($\etmiss$):
\item{\bullet} $\etmiss >50$ GeV,
\item{\bullet} no isolated leptons with $p_T(\ell)>15$ GeV,
\item{\bullet} number of jets $n(jet)\ge 4$,
\item{\bullet} at least one central jet $(|\eta (jet)|<1$) and
no jet within $30^o$ of $\etmissvec$.

\item{} This is a generic $\etmiss$ cut, {\it i.e.} it is not optimized
for various gluino masses. For instance, for very massive gluinos, a
substantially larger $\etmiss$ cut may be desirable to improve
signal to background rate.

\point Single charged lepton ($1\ell$):
\item{\bullet} exactly one isolated lepton with $p_T(\ell )>15$ GeV,
\item{\bullet} veto events with $60<M_T(\ell,\etmiss )<100$ GeV, to reduce
real $W$ background.

\point A pair of oppositely charged same-flavor leptons ($OS$):
\item{\bullet} two isolated leptons with $p_T(\ell )>15$ GeV,
\item{\bullet} $30^o<\Delta\phi (\ell^+\ell^- )<150^o$,
\item{\bullet} no jets, and
\item{\bullet} veto events with $80<M(\ell^+,\ell^- )<100$ GeV, to reduce
real $Z$ background.
\REF\bcpt{H. Baer, C. Chen, F. Paige and X. Tata, \prdj{49} (1994) 3283.}
\item{} These cuts are designed to extract possible signals from
slepton-pair production.\refmark\bcpt\

\point A pair of same-sign leptons ($SS$):
\item{\bullet} two same-charge isolated leptons with $p_T(\ell )>15$ GeV.

\REF\ssl{R. M. Barnett
{\it et. al.}, p. 159, in {\it High Energy Physics in the 1990's},
ed. S. Jensen, (World Scientific, 1989); H. Baer, X. Tata and J. Woodside,
\prdj{41} (1990) 906; R. M. Barnett, J.F. Gunion and H. Haber,
\plbj{315} (1993) 349.}
\item{} This is designed to extract gluino-pair cascade decay
events, by exploiting the Majorana nature of the gluino.\refmark\ssl\

\point Three leptons ($3\ell$):
\item{\bullet} three isolated leptons with $p_T(\ell_1 )>15$ GeV,
$p_T(\ell_2)>10$ GeV and $p_T(\ell_3)>8$ GeV,
\item{\bullet} veto events with $80<M(\ell^+,\ell^- )<100$ GeV, to reduce
real $WZ$ background,
\item{\bullet} (optional requirement of zero or one jet, or all jets
accompanying the event.)

\REF\bt{H. Baer and X. Tata, \prdj{47} (1993) 2739;
Lopez \etal, Ref.~[\lnwz];
H. Baer, C. Kao and X. Tata, \prdj{48} (1993) 5175.}
\item{} These cuts are designed to extract either gluino and squark cascade
decay events\refmark\bkt\ (with jets), or to extract clean trileptons from
$\cpmone\cntwo$ production\refmark\bt\ (with zero or one jet).

\point Four leptons ($4\ell$):\refmark\bkt\
\item{\bullet} four isolated leptons with the first three leptons as
in (5.) above, while in addition $p_T(\ell_4 )>8$ GeV.

\noindent
In addition, on occasion we picked up events containing five isolated
leptons. We do not list these relatively rare event cross sections
due to the considerable statistical uncertainty.

\endpage

\smallskip
\noindent{\bf 6. Numerical Results for Signal and Background}
\smallskip

The background cross section levels in the various channels after
cuts are given in Table~\xsecsbkgnd:
2720, $1.1\times 10^6$, $32$, $2$, $0.3$ (0.7), and $\sim 0$
in units of fb for the $\etmiss$, $1\ell$, $OS$, $SS$, $3\ell$
and $4\ell$ channels, respectively.  The backgrounds that were included
are: $W+jets$, $Z+jets$, $t\anti t$ ($\mt=170\gev$), $\wp\wm$,
and $\wpm Z$. The quoted rates include the $\tau$ mode decays of
the $\wpm$ and $Z$. The signal rates for each numbered case of
Fig.~\boundaries\ are given in Tables~\xsecs a--d, for the $\dil$, $\sdil$,
$\ns$ and $\sns$ models, respectively.

\FIG\etmissfig{The $\etmiss$ cross section (in fb) for
$p\anti p$ collisions at $\sqrt s=1.8\tev$ is plotted for all
the numbered scenarios in Fig.~\boundaries\ as a function
of $\mgl$; $\dil$, $\sdil$, $\ns$ and $\sns$ cases are indicated
by $d$, $D$, $m$ and $M$, respectively. The $\mgl$ mass value
can be used to identify the scenario within each such case.}

The $\etmiss +jets$ cross section after the above cuts is plotted
in Fig.~\etmissfig\ versus $\mgl$, for the various scenarios.
$\etmiss+jets$ events arise from many different sources
typically, including $\chitil\,\chitil$, $\gl\,\gl$, $\sq\,\gl$,
$\sq\,\sq$, $\stopi\,\stopi$, $\gl\,\chitil$, and $\sq\,\chitil$,
events, where all final state leptons are missed or soft.
All signals are below our calculated background of 2720 fb.
However, assuming that the background can be normalized
by independent measurements and Monte Carlo studies,
a $5\sigma$ effect for signal over background (given 1 $fb^{-1}$
of integrated luminosity) would allow a search to $\mgl\sim 300$ GeV.
This could be an overestimate given that at $\mgl\sim 300\gev$
the signal would only be $\sim 10\%$ of background, implying that
the latter would have to be normalized to better than 10\%.
However, it is also true that optimization of the $\etmiss$ cut
for these higher $\mgl$ values might improve the signal to background
ratio somewhat. The value of measuring the $\etmiss$ cross section
lies in the fact that it roughly scales with $\mgl$ in spite
of model differences, and different $\tanb$ values. Thus, if a $\etmiss +jets$
signal can be found, the size of the cross section will give an indication
of the sparticle masses being probed.

\FIG\rates{
We display the same-sign dilepton ($SS$) and
tri-lepton ($3\ell$) cross sections (fb) for $\sqrt s=1.8\tev$ $p\anti p$
collisions for each of the scenarios defined in Fig.~\boundaries.
No restrictions are placed on the number of associated jets.
The format is $[SS,3\ell]$ or $(SS,3\ell)$
where cases with square (curved) brackets have $\mhl<105\gev$ ($>105\gev$).
The dashed line indicates the approximate boundary
beyond which detection of a $SS$ or $3\ell$ signal would not be
possible for an integrated luminosity $L$ of $1\fbi$ or smaller.
The arrows, labelled by $\slepr$ and $\cpone$,
indicate the approximate $\mgl$ values corresponding to
$\mslepr=95\gev$ and $\mcpone=95\gev$, respectively.
The width of each arrow reflects the range
of $\mgl$ values obtained as $\tanb$ is varied at
fixed $\mslepr,\mcpone=95\gev$.
For $\mslepr<95\gev$, $\mcpone<95\gev$, $\mhl<105\gev$ detection of
$\slepr\slepr$, $\cpone\cmone$, $Z\hl$
production would be possible at LEP-200 with $L=500\pbi-1\fbi$.
}

The signal cross sections after cuts for the $1\ell$ sample are listed in
Tables~\xsecs a--d. These signals have an enormous background
from single $W$ production, in spite of the transverse mass cut we invoke.
Even with optimization of cuts ({\it e.g.} looking for events with
$M_T(\ell,\etmiss )>100$ GeV), detection of such a signal looks dubious, if
not hopeless.

The OS dilepton sample of signal events is suited for picking out
slepton-pair events. We see from Tables~\xsecs a--d that the
signal in this channel
can range to $\sim 80$ fb, although there is a substantial background
from $WW$ production (32 fb). We find signal larger
than or of order the background
in several cases: $\dilpvii$, $\dilmvii$, $\dilmviii$, $\nspi$, $\nspii$,
and $\snsmxii$.
These are more optimistic results than those given in Ref.~[\bcpt],
where slepton production was examined for more generic mass spectra.
Our larger rates are in part a reflection of the very light slepton
masses in the models considered here, and in part due
to the fact that numerous SUSY sources other than slepton pairs contribute to
the OS signal; these include mainly $\cpone\cmone$ pairs, but with smaller
contributions from $\chitil\,\chitil$, $\sq\,\chitil$,
$\gl\,\chitil$, $\stopi\,\stopi$ and $\sq\,\sq$ pairs.
Overall, however, observation of the OS dilepton signal
looks difficult for most of the cases examined.

The same-sign isolated dilepton signal has been advocated as a means
of searching for gluino-pair cascade decays, by exploiting the
Majorana nature of the gluino.\refmark\ssl\
We see from Tables~\xsecs a--d that the signal ranges from a fraction
of a fb to several hundred fb, while background is at the 2 fb level,
and arises (after cuts) mainly from $WZ$ production,
where one final state lepton is missed. In this case, many background events
should be relatively free of jet activity, while the signal may be rich in jets
if the SS events originate from strongly produced SUSY particles. We have
examined the sources of the SS events for the various $\dil$, $\sdil$,
$\ns$, and $\sns$
models, and have found that they arise from a variety of SUSY production
processes, including $\cpmone\cntwo\rta 3\ell+\etmiss$ events,
where one lepton is
missed, slepton and sneutrino production, $\gl\,\chitil$ and $\sq\,\chitil$
associated production events, as well as the expected $\sq\,\sq$, $\sq\,\gl$
and $\gl\,\gl$ events.
Hence, we expect the signal events to vary substantially in
topology and ``jetiness'', depending on the subprocess from which they arise.
Given an integrated luminosity of 1 $\fbi$, we take as an estimate
at least 5 such events to claim discovery. Examination of Tables~\xsecs a--d
then shows that scenarios with gluino masses of up to $\sim 300-350$ GeV may be
probed in this channel. We show in Fig.~\rates\ the SS dilepton signal
cross sections as the first entry in the brackets at each of the numbered
scenario points appearing in Fig.~\boundaries. (Except in cases
where overlap forced some slight repositioning, the \mgltb\
values for a particular point correspond to the
location of the lower left-hand corner of the bracket.)

Another promising event topology for the discovery of SUSY is events with
three isolated leptons. These events can arise from
$\cpmone\cntwo\rta 3\ell+\etmiss$ production,\refmark\bt\ in which case
they will be relatively
free of extra jet activity, or they can arise, for instance, from gluino and
squark production processes, where the cascade decays result in leptonically
decaying $\cpmone$ and $\cntwo$ states,\refmark\bkt\
in which case the trileptons will be
accompanied by substantial jet activity. We show the cross section for $3\ell$
events in Tables~\xsecs a--d for events containing just 0 or 1 jet,
or any number of
jets (in parenthesis). These cross sections range from a fraction of a fb,
to up to $168$ fb, for the cases examined. The background, listed
in Table~\xsecsbkgnd, is $0.34$ ($0.69$) fb. Again, assuming that at least
five events are needed for discovery (in 1 $\fbi$ of data), we see that
cases with gluino mass beyond 500 GeV may be probed
in the $\nsp$, $\nsm$, $\snsp$, $\dilp$, $\sdilp$ and $\sdilm$ models, while
in the $\dilm$ and $\snsm$ models discovery reach is restricted
to $\mgl\lsim 300\gev$. We list the trilepton
rates (for events with all jet multiplicities) in Fig.~\rates, as the second
entry in the bracketed figures.

By combining the discovery potential of both the
$SS$ and $3\ell$ signals, we have estimated the
region explorable by Tevatron collider experiments with $L=1\fbi$
of integrated luminosity. The boundary of this region for each model is
drawn on Fig.~\rates\ as the dashed line.
It is remarkable that so large a fraction of the model parameter
spaces will yield observable rates for these two new-physics signatures.
The most difficult scenarios to detect are those associated with
the $\dilm$ and $\snsm$ models. This is clearly a result of the
suppressed $\cpone\cntwo\rta 3\ell,SS$ mode (where
the SS events arise when one of the $\ell$'s is missed), as discussed earlier.
In addition,  in $\sq\,\gl$ \etc\ events
$\cpmone\cpmone\rta \ell\ell\snu\snu$
yields very soft leptons due to the generally
very small $\mcpmone-\msnu$ mass difference, which is much smaller
for these scenarios than any others.  As noted earlier, the difficulty
of probing the $\dilm$ and $\snsm$ scenarios is to be contrasted
with the situation for the $\nsm$ models, for which the inverted
mass hierarchy ($\mcpmone>\mslepl$ at large $\mgl$) allows for
much larger $3\ell$ and $SS$ rates.
Indeed, the $\nsm$ parameter space can be almost fully explored.

Finally, we list in Tables~\xsecs a--d the $4\ell$ event rates.
In a few cases, the rate
for $4\ell+\etmiss$ events can range up to the 30--60 fb level, giving again
a spectacular signature for SUSY. These events usually occur due to
events containing a $\cntwo$ pair, either from cascade decays, or from direct
production, followed by $\cntwo\rta \ell\bar{\ell}\cnone$ decay. We were unable
to generate any substantial background to this process. We note, however, that
it does not occur at a large rate for most of the scenarios considered,
and hence would not constitute an optimal discovery channel.

As noted earlier, extrapolation of these results to higher luminosity
is somewhat dangerous, but we allow ourselves a few very approximate
statements based on the $3\ell$ mode which has the lowest background
rate.  For an integrated luminosity of $L=30\fbi$
(three years running at $10\fbi$ per year as proposed
in some versions of a future Tevatron upgrade
\Ref\jackson{G. Jackson, talk at SUSY-94, Ann Arbor, MI.}), the $3\ell$
background rate (assuming no additional sources of background
become important at high luminosity, \eg\ from multiple
interactions per crossing) is obtained from Table~\xsecsbkgnd;
we find a background rate of 21 events for the `all-jets' case.
An examination of the signal rates in Tables~\xsecs a--d shows that
many of the numbered scenarios that are unobservable for $L=1\fbi$
would then become observable.  The increase in parameter space coverage
would be especially dramatic for the $\dilm$ and $\snsm$ models.
Adopting a 5$\sigma$ criterion (\ie\ 23 or more signal events),
$\mgl$ values as high as $\sim 500\gev$ would yield detectable
signals for the $\dilm$ and $\snsm$ models.  For the other models,
we have not studied scenarios with high enough $\mgl$ to establish
a meaningful estimate of how much higher in $\mgl$ one can
go with $L=30\fbi$. However, those scenarios we have studied suggest
that discovery reach would probably be extended out to $\mgl\sim 600-700\gev$.

\smallskip
\noindent{\bf 7. Comparison of Tevatron and LEP-II}
\smallskip

An interesting question is the extent to which LEP-II will be able
to explore the \mgltb\ parameter spaces of the various models
considered, and how the discovery reach of LEP-II compares
to that of the Tevatron. For the models being considered
the discovery reach of LEP-II is determined primarily by
the $\ep\em\rta\slepr\slepr$, $\ep\em\rta\cpone\cmone$,
and $\ep\em\rta Z\hl$ production
processes, since it is the $\slepr$, $\cpmone$ and $\hl$ that are the
lightest observable particles in these models.

\FIG\lep{
We plot the contours for $\mslepr=83,95\gev$ (dots),
$\mcpone=83,95\gev$ (dashes) and $\mhl=81,105\gev$ (dotdash)
within  the allowed \mgltb\
parameter space for the eight models illustrated in Fig.~\boundaries.
The lower and upper mass values for each particle type
represent rough upper limits for the mass reach at LEP-II
with $\sqrt s=176,200\gev$, respectively.
In the $\sdilp$ case, $\mslepr\geq 83\gev$ for all allowed $\mgl,\tanb$,
and only the $\mslepr=95\gev$ contour appears.
In the $\nsm$ case, only the $\mslepr=83\gev$ contour (barely) appears
since $\mslepr\leq 95\gev$ throughout all of the allowed parameter space.
In the $\dilp$, $\sdilp$ and $\snsp$ windows, the $\mhl=81\gev$ contour
is barely visible in the low-$\tanb$, low-$\mgl$ parameter space
corner. In the $\nsm$ model,
$\mhl\leq 105\gev$ for all allowed $\mgl,\tanb$ values,
and only the $\mhl=81\gev$ contour appears.
}

Let us first consider an optimistic scenario with $\sqrt s=200\gev$
and a two-year accumulated luminosity of $1\fbi$ (corresponding
to $500\pbi$ per year integrated over two years).
Then, roughly speaking, $\slepr\slepr,\cpone\cmone$
production will probably be detectable for $\mslepr,\mcpmone\lsim 95\gev$,
while $Z\hl$ will probably be detectable for $\mhl\lsim 105\gev$
(recalling that the $\hl$ has $ZZ$ coupling that is close to
full strength given the large $\mha$ values required in our scenarios).
The contours for these $\slepr$, $\cpmone$ and $\hl$ mass values
are given in Fig.~\lep, for all eight of the models illustrated
in Fig.~\boundaries.  Specific $\mslepr$, $\mcpmone$ and $\mhl$
values for all the numbered scenarios have been
listed in Tables~\massest a--d.
As a further aid to comparing LEP-II and Tevatron results we have also
indicated rough LEP-II discovery potential for the $\slepr\slepr$,
$\cpone\cmone$ and $Z\hl$ channels in Fig.~\rates\ as described below.

For $\slepr\slepr$-pair production, we have used an arrow
in each model window to indicate the approximate upper limit in $\mgl$
for which slepton-pair production will
be observable at LEP-200. The width of the arrow characterizes
the variation in $\mgl$ associated with fixed $\mslepr=95\gev$
apparent in Fig.~\lep.
In the $\ns$, $\dil$, $\sns$ and $\sdil$ models, the values of $\mgl$
for which $\mslepr=95\gev$ fall in the ranges [636,665],
[351,370], [469,493], [264,279] GeV, respectively.
$\mslepr$ and, hence, these ranges are independent of the sign of $\mu$.
If only $\mslepr=92\gev$ could be probed, the corresponding
$\mgl$ ranges become [611,641], [337,357],
[450,475] and [253,268] GeV, respectively. Note that $\slepr\slepr$
detection will be possible for essentially
all of the $\nsm$ allowed parameter space,
and for almost none of $\sdilp$ parameter space.

A similar procedure is followed in the case of
$\cpone\cmone$-pair production.
For $\sqrt s=200\gev$, we adopt $\mcpmone=95\gev$ as the discovery boundary.
For a given model, the $\mgl$ value corresponding to $\mcpmone=95\gev$
depends upon $\tanb$ as indicated in Fig.~\lep. Note
that all the $\mu>0$ contours are in roughly the same $\mgl$
mass range, as are all the $\mu<0$ contours. A rough summary is that
the $\mgl$ ranges corresponding to $\mcpmone=95\gev$ fall within the bands:
$[381,434]\gev$ for $\dilp$, $\sdilp$, $\nsp$ and $\snsp$; and
$[282,385]\gev$ for $\dilm$, $\sdilm$, $\nsm$ and $\snsm$.
Thus, discovery of a chargino of mass $\mcpmone=95\gev$ on the average
probes significantly larger $\mgl$ values for $\mu>0$ than for $\mu<0$.
The arrows labelled by $\cpone$ in Fig.~\rates\ reflect the above ranges.
This dependence on the sign of $\mu$ is also evident in Fig.~\masses.
There, the $\mcpmone/\mgl$ mass bands for $\mu>0$ rise towards
the large $\mgl$ asymptotic limit as $\mgl$ increases, whereas for
$\mu<0$ the bands lie above the large-$\mgl$ limit,
and always somewhat above the $\mu>0$ band. Note that in the $\nsm$
model, $\cpone\cmone$ pair production  will
be detectable for almost none of the allowed parameter space,
in sharp contrast to the guaranteed discovery of $\slepr\slepr$ pairs
for this model.

In comparison to the $\slepr\slepr$ discovery limits quoted previously,
we see that $\cpone\cmone$ pair production does not reach to as large
$\mgl$ in the $\ns$ and $\sns$ cases, whereas in the $\dil$ and $\sdil$
cases $\mgl$ values probed are larger (comparable) for $\mu>0$ ($\mu<0$).
These trends become especially apparent by comparing the
$\mslepr$ and $\mcpmone$ contours in Fig.~\lep.

With regard to the $Z\hl$ mode at LEP-II,
those numbered scenarios for which $\mhl<105$ ($>105$) GeV
are surrounded by square (rounded) brackets in Fig.~\rates.
However, the precise boundary of detectability is very sensitive
to precise luminosity, detector efficiencies and machine
energy. As an indication of this, we note that
if only $\mhl\leq 100\gev$ could be probed,
then $Z\hl$ production would {\it not} be detectable for
the numbered scenarios $\dilpvi$, $\dilpii$, $\dilpv$, $\dilmvi$,
$\dilmiii$, $\nspii$, $\nspv$, $\nsmvi$, $\nsmiii$, $\snsmiii$,
$\snsmii$ and $\sdilmviii$ (\ie\ their square brackets would become rounded
brackets in Fig.~\rates). These scenarios, for which $\mhl\gsim 100\gev$,
are mostly those with larger values of $\tanb$ --- recall that $\mhl$
is smallest, even after radiative corrections, for $\tanb$ near 1.

The $Z\hl$ detection boundaries are more clearly indicated, however,
in Fig.~\lep. We note that the $\mhl=105\gev$ contour is absent
from the $\nsm$ window of Fig.~\lep c,
since in the $\nsm$ model $\mhl<105\gev$
for all of allowed parameter space and $Z\hl$
detection will always be possible at LEP-II (for $\sqrt s=200\gev$).

Of course, detection of the $\hl$ at the Tevatron may also be possible
in the $W\hl$ associated production channel with
$\hl\rta b\anti b$ decay, provided there is adequate
efficiency and purity for $b$ tagging.\Ref\willenetal{A. Stange,
W. Marciano, and S. Willenbrock, preprint ILL-TH-94-8.}\
The upper $\mhl$ limits for
which this will be possible are luminosity dependent.
For $L=1,10\fbi$ the upper limit is $\mhl\lsim 60,75\gev$.
As can be seen in Tables \massest a--d,
rather few of our scenarios have masses below $75\gev$, and only one
has mass below $60\gev$. Even $L=100\fbi$ at the Tevatron would only
allow one to probe $\mhl$ masses up to $\sim 95\gev$.

The bottom line is clear: LEP-II at $\sqrt s=200\gev$ and full luminosity
can generally probe much the same parameter space as can the Tevatron.
The coverage is comparable for the $\dilp$ and $\snsp$ models, somewhat greater
in the case of the $\dilm$, $\nsp$, $\nsm$ and $\snsm$ models, and somewhat
less in the case of the $\sdilp$ and $\sdilm$ models.
However, generally speaking, the Tevatron is sensitive to a much broader
set of SUSY particles than is LEP-II, although LEP-II does have
sensitivity to both slepton pairs and chargino pairs for $\mgl$
below a model-dependent value.
Obviously, there is substantial complementarity between the two machines.

As already noted, these conclusions are significantly altered if
LEP-II only reaches, say, $\sqrt s=176\gev$. Because
the $\hl$ often has mass of the order of $100\gev$, for the chosen $\mt$
and typical $\stop$ masses predicted in the models being considered,
the extent to which $Z\hl$ discovery will probe
the allowed parameter spaces of Fig.~\boundaries\
is {\it extremely} sensitive to
the exact $\sqrt s$ and luminosity that will be achieved at LEP-II.
For $\sqrt s=176\gev$, $Z\hl$ detection will at best
only be possible for $\mhl\lsim 81\gev$ (comparable
to the $L=10\fbi$ Tevatron reach).  The $\mhl=81\gev$ contour
for each model is given in Fig.~\lep.  Note how much less of
parameter space would allow $Z\hl$ detection.
Only for numbered scenarios with low $\tanb$ values would $Z\hl$
detection be possible, which amounts to only some 6 or 7 of the scenarios ---
see the $\mhl$ masses in Tables~\massest a--d.

At $\sqrt s=176\gev$,
slepton-pair production would be viable only for $\mslepr\lsim 80-83\gev$,
which
corresponds to $\mgl\lsim [504,568]$, [277,316], [373,421], [210,238] GeV
(where the ranges now include both $\tanb$ variation and a
$80-83\gev$ range of possible mass accessibility)
for the $\ns$, $\dil$, $\sns$ and $\sdil$ models, respectively, \ie\ some
$70-100\gev$ below the $\mgl$ values quoted earlier for $\sqrt s=200\gev$.
Chargino-pair production would be viable for $\mcpmone\lsim 83\gev$,
which corresponds to the rough ranges $\mgl\lsim [346,405]$ for $\mu>0$ cases,
and $\mgl\lsim [233,344]$ for $\mu<0$ cases.
More precise limits as a function of model and $\tanb$
for $\sqrt s=176\gev$ are reflected by the
$\mslepr=83\gev$ and $\mcpmone=83\gev$ contours given
in Fig.~\lep. The bottom line is clear.
The slepton-pair and chargino-pair modes at LEP-II would be
generally competitive with the Tevatron for the $\dilm$ and $\snsm$
models, but the Tevatron would provide a signal for SUSY over
more of parameter space for the $\dilp$, $\sdilp$, $\sdilm$,
$\nsp$, $\nsm$, and $\snsp$ models.

\smallskip
\noindent{\bf 8. Conclusion}
\smallskip

In this paper, we have explored the phenomenology of
the gauge-coupling-unified Minimal Supersymmetric Model
employing renormalization group evolution of
superstring or supergravity motivated
unification-scale boundary conditions
(of a rather universal and attractive nature)
for the soft-supersymmetry-breaking parameters.
The models considered were the minimal-supergravity (or no-scale) and
the dilaton-like models. At a theoretical level,
the source of supersymmetry breaking
and details of the Kahler potential, and so forth, are fairly
different for the two model classes, and even the dilaton-like
boundary conditions themselves apply for a wide variety of
physics as contained in the continuous range of possible values for
the goldstino angle characterizing
the relative importance of moduli vs. pure dilaton supersymmetry
breaking. Despite these theoretical differences, the boundary
conditions are sufficiently similar (indeed, identical
within the dilaton-like class) that there is a broad similarity
of the basic phenomenology of these models, deriving from
the presence of relatively light sleptons in all cases.
In fact, we have seen that when the uncertainty
associated with the question of whether unification should be required
at $\mgut\sim 2\times 10^{16}\gev$ or at $M_S\sim 10^{18}\gev$
is taken into account, the overlap between the mass spectra
and resulting phenomenology of the minimal-supergravity and dilaton-like
models can be quite substantial.  Nonetheless,
we have also seen that seemingly small shifts in mass spectra
can cause substantial shifts in allowed decay modes and the
consequent visibility of crucial detection channels.

Overall, the most remarkable feature of our results is the prediction
(summarized in Fig.~\rates) that these classes of models
can be probed by the
existing Tevatron (with $L=1000\pbi$) over such a large portion of
the allowed parameter spaces. In terms of the two parameters $\mgl$ and
$\tanb$, we find that even the most difficult models, namely the
$\mu<0$ dilaton-equivalent ($\dilm$) and superstring-scale-unified
minimal-supergravity models ($\snsm$), yield observable
tri-lepton and same-sign-lepton signals for $\mgl\lsim 300\gev$
(for all $\tanb$). For the $\mu>0$, $\dilp$ and $\snsp$ models
the $3\ell$ and $SS$ signals reach
observable levels for $\mgl$ values as high as $600\gev$ at low $\tanb$
(as preferred if relatively precise Yukawa coupling constant unification
is demanded). The $3\ell$ signal reaches an observable
level for $\mgl$ values up to $450-520\gev$ in
the $\sdil^{\pm}$ models.
The GUT-scale-unified minimal-supergravity $\mu>0$ ($\nsp$) model
can be probed for $\mgl\lsim 420-480\gev$.  This represents somewhat
more than half of the allowed parameter space given that $\mgl$
has an upper bound (deriving from the requirement of a neutral LSP)
in this model of about $700\gev$ (or lower at high $\tanb$).
Meanwhile, the $\nsm$ model can be probed over nearly all of the (rather
restricted) parameter space.

Thus, there is cause for optimism that the scheduled main-injector
upgrade of the Tevatron will reveal evidence for supersymmetry.
However, there is no guarantee.  Aside
from the regions of parameter space for the models discussed here
that lie beyond the reach of the Tevatron, there are also the
(still more model dependent) moduli-dominated scenarios in which
all sfermion masses are generically expected to be larger than the
gluino mass.  These, as discussed in Ref.~[\bkt] and
\REF\bdknt{H. Baer, M. Drees, C. Kao, M. Nojiri and X. Tata,
FSU-HEP-940311 (1994).}
Ref.~[\bdknt], are more difficult to probe without
a collider of significantly larger energy.  Larger energies
and luminosities could also improve observability of the
$\dil$, $\sdil$, $\ns$ and $\sns$ models.  This is under investigation.
Here, we note that the predicted $SS$ and $3\ell$ background rates
imply that the discovery reach is not far from being background limited.
Thus, simply increasing luminosity may not yield as large an improvement
as would otherwise be the case.
Increasing the energy may be more advantageous due to increased signal
rates, although background rates will also go up and new backgrounds
can arise.

Finally, we have noted that for some models
LEP-II with $\sqrt s=200\gev$ and integrated
luminosity of $500\pbi-1\fbi$ will be able to detect $\slepr\slepr$
and $\cpone\cmone$ pair production, and $Z\hl$ associated Higgs production,
over more of \mgltb\ parameter space than that for which a SUSY signal will
be seen at the Tevatron with $L=1\fbi$ of integrated luminosity.
However, in other models LEP-200 will probe less of parameter space.
Further, the relative comparison between the two machines is very
dependent upon the precise energy reached by LEP-II and on
whether further luminosity upgrades for the Tevatron are implemented.
In general, the two machines are quite complementary, with
the Tevatron being sensitive to a broader range of SUSY particle types
in those regions of parameter space for which SUSY detection is possible.

\bigskip
\REF\klmw{T. Kamon, J.L. Lopez, P. McIntyre, and J.T. White,
CTP-TAMU-19/94.}
\noindent{\it Note Added:} As we were completing this manuscript we
received a paper\refmark\klmw\ which also addresses the search
for minimal supergravity at the Tevatron and Di-Tevatron.
\smallskip

\smallskip\noindent{\bf 9. Acknowledgements}
\smallskip
We thank X. Tata for discussions.
This work has been supported in part by Department of Energy
grant \#DE-FG05-87ER40319, \#DE-FG03-91ER40674, and by
Texas National Research Laboratory grant \#RGFY-93-330. In addition, the work
of HB was supported by a TNRLC SSC fellowship.

\smallskip
\refout

 \pageinsert
 \titlestyle{\tenpoint
 Table \massest a: A tabulation of supersymmetric particle masses
for the $\dil$ scenarios delineated in Fig.~\boundaries a.
}
 \smallskip
\def\tstrut{\vrule height 12pt depth 4pt width 0pt}
 \thicksize=0pt
 \hrule \vskip .04in \hrule
 \begintable
 Scenario | $\mgl$ | $\tanb$ | $\mhl$ | $\mha$ | $\mcnone$ | $\mcntwo$ |
$\mcpone$ | $\mslepl$ |   $\mslepr$ | $\msnu$ | $\msq$ | $\mstopi$ \cr
$\dilpvii$ | 282 |  2.0 | 75.3 | 240 |
23.4 | 59.1 | 47.7 | 96.8 | 75.5 | 74.4 | 252 | 158 \nr
$\dilpii$ | 295 | 9.0  | 101 | 180 |
35.5 | 66.3 | 62.4 | 105  | 82.8 | 68.7 | 264 | 178 \nr
$\dilpvi$ | 310 | 15.0 | 103 | 180 |
39.9 | 72.5 | 70.2 | 109  | 85.9 | 74.4 | 277 | 188 \nr
$\dilpi$ | 346 | 3.2  | 93.6 | 250 |
40.4 | 79.2 | 73.5 | 118  | 91.8 | 93.0 | 310 | 195 \nr
$\dilpv$ | 431 | 4.5  | 104 | 300 |
58.4 | 109  | 107  | 144  | 111  | 122  | 386 | 250 \nr
$\dilpviii$ | 435 | 8.0  | 108 | 287 |
60.8 | 113  | 112  | 146  | 112  | 123  | 390 | 258 \nr
$\dilpiii$ | 503 | 5.0  | 108 | 350 |
71.3 | 134  | 133  | 166  | 127  | 147  | 450 | 297 \nr
$\dilpiv$ | 609 | 2.0  | 95.6 | 550 |
87.9 | 169  | 168  | 197  | 150  | 187  | 545 | 357 \nr
$\dilmvii$ | 232 | 2.0  | 58.4 | 190 |
37.1 | 83.5 | 83.3 | 82.3 | 65.0 | 54.1 | 207 | 215 \nr
$\dilmviii$ | 242 | 3.2  | 76.9 | 162 |
37.8 | 73.8 | 76.0 | 88.1 | 70.2 | 50.0 | 217 | 192 \nr
$\dilmii$ | 295 | 9.0  | 98.4 | 180 |
42.6 | 76.5 | 77.4 | 105  | 82.8 | 68.7 | 264 | 195 \nr
$\dilmix$ | 301 | 2.2  | 69.0 | 244 |
47.3 | 100  | 100  | 103  | 80.2 | 79.8 | 269 | 242 \nr
$\dilmvi$ | 310 | 15.0 | 101  | 180 |
43.9 | 78.7 | 79.0 | 109  | 60.3 | 74.4 | 277 | 198 \nr
$\dilmi$ | 346 | 3.2  | 86.0 | 250 |
53.4 | 106  | 106  | 118  | 91.8 | 93.0 | 310 | 246 \nr
$\dilmv$ | 431 | 4.5  | 98.4 | 300 |
65.6 | 128  | 128  | 144  | 111  | 122  | 386 | 285 \nr
$\dilmiii$ | 503 | 5.0  | 103  | 350 |
76.9 | 150  | 151  | 166  | 127  | 147  | 450 | 329 \nr
$\dilmiv$ | 609 | 2.0  | 81.3 | 550 |
95.4 | 193  | 193  | 197  | 150  | 187  | 545 | 436
\endtable
 \hrule \vskip .04in \hrule
\bigskip
 \titlestyle{\tenpoint
 Table \massest b: A tabulation of supersymmetric particle masses
for the $\sdil$ scenarios delineated in Fig.~\boundaries b.
}
 \smallskip
\def\tstrut{\vrule height 12pt depth 4pt width 0pt}
 \thicksize=0pt
 \hrule \vskip .04in \hrule
 \begintable
 Scenario | $\mgl$ | $\tanb$ | $\mhl$ | $\mha$ | $\mcnone$ | $\mcntwo$ |
$\mcpone$ | $\mslepl$ |   $\mslepr$ | $\msnu$ | $\msq$ | $\mstopi$ \cr
$\sdilpiii$ | 471 | 15.0 | 112  | 357 |
67.4 | 129  | 129  | 193  | 157  | 175  | 464 | 289 \nr
$\sdilpvii$ | 503 | 5.0  | 110 | 424 |
71.1 | 136  | 135  | 205  | 166  | 190  | 496 | 303 \nr
$\sdilpviii$ | 510 | 2.0  | 93.7| 542 |
70.5 | 136  | 135  | 206  | 167  | 196  | 503 | 308 \nr
$\sdilmv$ | 471 | 15.0 | 111  | 357 |
69.1 | 134  | 134  | 193  | 157  | 176  | 464 | 301 \nr
$\sdilmviii$ | 503 | 5.0  | 105  | 424 |
75.4 | 149  | 149  | 205  | 166  | 190  | 496 | 339 \nr
$\sdilmix$ | 510 | 2.0  | 78.5 | 542 |
78.0 | 159  | 159  | 206  | 167  | 196  | 503 | 392
\endtable
 \hrule \vskip .04in \hrule
\vfill
 \endinsert

 \pageinsert
 \titlestyle{\tenpoint
 Table \massest c: A tabulation of supersymmetric particle masses
for the $\ns$ scenarios delineated in Fig.~\boundaries c.
}
 \smallskip
\def\tstrut{\vrule height 12pt depth 4pt width 0pt}
 \thicksize=0pt
 \hrule \vskip .04in \hrule
 \begintable
 Scenario | $\mgl$ | $\tanb$ | $\mhl$ | $\mha$ |
$\mcnone$ | $\mcntwo$ | $\mcpone$ | $\mslepl$ |
  $\mslepr$ | $\msnu$ | $\msq$ | $\mstopi$ \cr
$\nspi$ |  296 | 2.2 | 78.3 | 220 |
26.0 | 62.6 | 51.3 | 82.5 | 52.9 | 51.0 | 257 | 171  \nr
$\nspii$ |  324 | 9.0 | 102  | 180 |
40.3 | 74.4 | 70.7 | 92.6 | 60.9 | 48.2 | 281 | 206  \nr
$\nspviii$ |  424 | 8.0 | 106  | 252 |
58.4 | 108  | 106  | 115  | 71.0 | 83.6 | 370 | 266  \nr
$\nspv$ |  471 | 4.5 | 104  | 300 |
65.2 | 122  | 119  | 125  | 75.2 | 99.0 | 410 | 290  \nr
$\nspvii$ |  491 | 2.2 | 92.0 | 387 |
67.0 | 127  | 125  | 127  | 74.3 | 110  | 428 | 285  \nr
$\nspvi$ |  492 | 7.0 | 108  | 300 |
69.9 | 130  | 129  | 130  | 78.1 | 104  | 429 | 309  \nr
$\nspiii$ |  550 | 5.0 | 108  | 350 |
79.2 | 149  | 148  | 143  | 84.3 | 121  | 479 | 344  \nr
$\nspiv$ |  605 | 2.0 | 82.5 | 510  |
87.3 | 167.4| 166  | 154.1| 87.4 | 141.1| 528 | 388  \nr
$\nsmi$ |  296 | 2.2 | 68.0 | 220 |
47.0 | 98.5 | 98.7 | 82.5 | 52.9 | 51.0 | 257 | 241  \nr
$\nsmii$ |  324 | 9.0 | 99.1 | 180 |
47.1 | 84.6 | 85.5 | 92.6 | 60.9 | 48.2 | 281 | 221  \nr
$\nsmiv$ |  368 | 2.0 | 69.4 | 300 |
58.0 | 121  | 120  | 98.1 | 59.5 | 76.1 | 321 | 288  \nr
$\nsmv$ |  373 | 4.5 | 94.5 | 230 |
56.7 | 108  | 109  | 103  | 64.7 | 68.7 | 325 | 260  \nr
$\nsmvi$ |  400 | 7.0 | 101  | 238 |
59.8 | 112  | 113  | 109  | 68.3 | 76.2 | 349 | 270  \nr
$\nsmvii$ |  450 | 3.0 | 89.1 | 311 |
69.8 | 139  | 139  | 119  | 71.4 | 95.1 | 392 | 319  \nr
$\nsmiii$ |  477 | 5.0 | 101  | 300 |
72.9 | 141  | 141  | 127  | 76.1 | 100  | 416 | 324
\endtable
 \hrule \vskip .04in \hrule
\bigskip
 \titlestyle{\tenpoint
 Table \massest d: A tabulation of supersymmetric particle masses
for the $\sns$ scenarios delineated in Fig.~\boundaries d.
}
 \smallskip
\def\tstrut{\vrule height 12pt depth 4pt width 0pt}
 \thicksize=0pt
 \hrule \vskip .04in \hrule
 \begintable
 Scenario | $\mgl$ | $\tanb$ | $\mhl$ | $\mha$ |
$\mcnone$ | $\mcntwo$ | $\mcpone$ | $\mslepl$ |
  $\mslepr$ | $\msnu$ | $\msq$ | $\mstopi$ \cr
$\snspix$ |  390 | 8.0 | 106  | 274 |
52.9 | 98.8 | 97.4 | 124 | 82.1 | 96.2 | 370 | 253 \nr
$\snspvii$ |  570 | 2.2 | 97.6 | 529 |
81.0 | 156 | 155 | 173 | 108 | 160 | 541 | 355 \nr
$\snspviii$ |  591 | 4.5 | 110  | 453 |
85.5 | 164 | 164 | 180 | 114 | 163 | 561 | 386 \nr
$\snsmxii$ |  290 | 9.0 | 98.6 | 192 |
41.5 | 76.2 | 77.0 | 97.5 | 67.6 | 57.1 | 275 | 209 \nr
$\snsmxiii$ |  300 | 3.0 | 81.7 | 234 |
45.9 | 92.9 | 93.4 | 98.2 | 66.5 | 67.2 | 284 | 239 \nr
$\snsmxiv$ |  300 | 5.0 | 93.4 | 212 |
44.6 | 85.0 | 86.0 | 99.6 | 68.3 | 63.4 | 285 | 223 \nr
$\snsmxi$ |  301 | 6.0 | 95.9 |  208 |
44.2 | 83.2 | 84.1 | 100 | 68.7 | 62.9 | 285 | 220
\endtable
 \hrule \vskip .04in \hrule
 \vfill
\endinsert

\pageinsert
 \titlestyle{\tenpoint
 Table \brs a: A tabulation of some important branching ratios
for the $\dil$ scenarios delineated in Fig.~\boundaries a.
Results are quoted for a single $l$ or $\nu$ type (\eg\ $l=e$);
for the $\cntwo$ particle-antiparticle and antiparticle-particle
channels are summed together.
}
 \smallskip
\def\tstrut{\vrule height 12pt depth 4pt width 0pt}
 \thicksize=0pt
 \hrule \vskip .04in \hrule
\begintable
 \    | \multispan{2} \tstrut\hfil $\cpone\rta$ \hfil |
        \multispan{3} \tstrut\hfil $\cntwo\rta$ \hfil |
        \multispan{3} \tstrut\hfil $\slepl\rta$ \hfil |
        \multispan{2} \tstrut\hfil $\snu\rta$\hfil  \nr
 Scenario | $\snu l$ & $l$-3-body |
            $l\slepr$ & $\nu\snu$ & $l$-3-body |
            $\cpone\nu$ & $l \cnone$ & $l\cntwo$ |
            $\cpone l$ & $\cntwo\nu$ \cr
$\dilpvii$ |
0.0 & 0.21 | 0.0 & 0.0 & 0.19 |
0.64 & $\sim 0$ & 0.36 |  0.48 & 0.025     \cr
$\dilpii$ |
0.0 & 0.29 | 0.0 & 0.0 & 0.046 |
0.54 & 0.050 & 0.42 |  0.16 & 0.005     \cr
$\dilpvi$ |
0.0 & 0.30 | 0.0 & 0.0 & 0.041 |
0.51 & 0.094 & 0.39 |  0.085 & 0.004     \cr
$\dilpi$ |
0.0 & 0.21 | 0.0 & 0.0 & 0.15 |
0.59 & 0.037 & 0.37 |  0.37 & 0.046     \cr
$\dilpv$ |
0.0 & 0.21 | 0.0 & 0.0 & 0.16 |
0.52 & 0.16 & 0.32 |   0.26 & 0.063    \cr
$\dilpviii$ |
0.0 & 0.23 | 0.004 & 0.0 & 0.04 |
0.48 & 0.22     & 0.30 |  0.18 & 0.049      \cr
$\dilpiii$ |
0.0 & 0.20 | 0.089 & 0.0 & 0.044 |
0.46 & 0.27 & 0.27 |   0.20 & 0.06    \cr
$\dilpiv$ |
0.0 & 0.14 | 0.096 & 0.0 & 0.069 |
0.40 & 0.38 & 0.22 |   0.21 & 0.077    \cr
$\dilmvii$ |
0.33 & $\sim 0$ | 0.006 & 0.32 & $\sim0$ |
0.0 & 1.0 & 0.0 |       0.0 & 0.0    \cr
$\dilmviii$ |
0.33 & 0.001 | $\sim 0$ & 0.33 & 0.001 |
0.17 & 0.70 & 0.13 | 0.0     & 0.0   \cr
$\dilmii$ |
0.33 & $\sim0$ | 0.0 & 0.33 & 0.001 |
0.29 & 0.31 & 0.40 |    0.0 & 0.0    \cr
$\dilmix$ |
0.32 & 0.01 | 0.004 & 0.32 & $\sim 0$ |
0.02 & 0.97 & 0.01 |       0.0 & 0.0    \cr
$\dilmvi$ |
0.33 & $\sim 0$ | 0.11 & 0.22 & 0.001 |
0.43 & 0.24 & 0.33 |    0.0 & 0.0    \cr
$\dilmi$ |
0.32 & 0.007 | $\sim 0$ & 0.32 & 0.007 |
0.16 & 0.74 & 0.096 |   0.0 & 0.0    \cr
$\dilmv$ |
0.31 & 0.014 | $\sim 0$ & 0.30 & 0.026 |
0.22 & 0.65 & 0.13 |    0.0 & 0.0    \cr
$\dilmiii$ |
0.26 & 0.036 | 0.002 & 0.23 & 0.076 |
0.19 & 0.70 & 0.11 |    0.0 & 0.0    \cr
$\dilmiv$ |
0.036 & 0.022 | $\sim 0$ & 0.020 & 0.015 |
0.011 & 0.98 & 0.005 |  0.0 & 0.0
\endtable
 \hrule \vskip .04in \hrule
\bigskip
 \titlestyle{\tenpoint
 Table \brs b: A tabulation of some important branching ratios
for the $\sdil$ scenarios delineated in Fig.~\boundaries b.
Results are quoted for a single $l$ or $\nu$ type (\eg\ $l=e$);
for the $\cntwo$ particle-antiparticle and antiparticle-particle
channels are summed together.
}
 \smallskip
\def\tstrut{\vrule height 12pt depth 4pt width 0pt}
 \thicksize=0pt
 \hrule \vskip .04in \hrule
\begintable
 \    | \multispan{2} \tstrut\hfil $\cpone\rta$ \hfil |
        \multispan{3} \tstrut\hfil $\cntwo\rta$ \hfil |
        \multispan{3} \tstrut\hfil $\slepl\rta$ \hfil |
        \multispan{2} \tstrut\hfil $\snu\rta$\hfil  \nr
 Scenario | $\snu l$ & $l$-3-body |
            $l\slepr$ & $\nu\snu$ & $l$-3-body |
            $\cpone\nu$ & $l \cnone$ & $l\cntwo$ |
            $\cpone l$ & $\cntwo\nu$ \cr
$\sdilpiii$ |
0.0 & 0.16 | 0.0 & 0.0 & 0.09 |
0.51 & 0.18 & 0.30 |  0.49 & 0.20      \cr
$\sdilpvii$ |
0.0 & 0.14 | 0.0 & 0.0 & 0.10  |
0.53 & 0.15  & 0.31 |  0.50 & 0.20      \cr
$\sdilpviii$ |
0.0 & 0.14 | 0.0 & 0.0 & 0.13  |
0.55 & 0.13  & 0.32 |  0.50  & 0.20      \cr
$\sdilmv$ |
0.0  &  0.19    | 0.0   & 0.0  & 0.10    |
0.49 & 0.23 & 0.28 |       0.48  & 0.21    \cr
$\sdilmviii$ |
0.0 & 0.17 | 0.0 & 0.0 & 0.13 |
0.47 & 0.28 & 0.25 | 0.46      & 0.22   \cr
$\sdilmix$ |
0.0  & 0.11 | 0.0   & 0.0  & 0.004    |
0.43 & 0.37 & 0.20 |      0.43 & 0.20
\endtable
 \hrule \vskip .04in \hrule
 \vfill
\endinsert

\pageinsert
 \titlestyle{\tenpoint
 Table \brs c: A tabulation of some important branching ratios
for the $\ns$ scenarios delineated in Fig.~\boundaries c.
Results are quoted for a single $l$ or $\nu$ type (\eg\ $l=e$);
for the $\cntwo$ particle-antiparticle and antiparticle-particle
channels are summed together.
}
 \smallskip
\def\tstrut{\vrule height 12pt depth 4pt width 0pt}
 \thicksize=0pt
 \hrule \vskip .04in \hrule
 \begintable
 \    | \multispan{3} \tstrut\hfil $\cpone\rta$ \hfil |
        \multispan{4} \tstrut\hfil $\cntwo\rta$ \hfil |
        \multispan{3} \tstrut\hfil $\slepl\rta$ \hfil \nr
 Scenario | $\snu l$ & $\slepl \nu$ & $l$-3-body |
            $l\slepr$ & $l\slepl$ & $\nu\snu$ & $l$-3-body |
            $\cpone\nu$ & $l \cnone$ & $l\cntwo$ \cr
$\nspi$ |
0.27  & 0.0  & 0.048 | 0.18  & 0.0  & 0.15  & $\sim 0$ | 0.71 & $\sim 0$ & 0.29
\cr
$\nspii$ |
0.33  & 0.0  & $\sim 0$ | 0.049  & 0.0  & 0.28  & $\sim 0$ | 0.54 & 0.12 & 0.34
\cr
$\nspviii$ |
0.33 & 0.0   & 0.001  | 0.048 & 0.0  & 0.28  & $\sim 0$ | 0.25 & 0.64 & 0.11
\cr
$\nspv$ |
0.32 & 0.0   & 0.002    | 0.058 & 0.0  & 0.27  & $\sim 0$ | 0.13  & 0.84 & 0.03
\cr
$\nspvii$ |
0.31 & 0.0   & 0.006  | 0.080 & 0.0  & 0.24  & $\sim 0$ | 0.04 & 0.96 & $\sim0$
\cr
$\nspvi$ |
0.32 & 0.0   & 0.002    | 0.058 & 0.0  & 0.27  & $\sim 0$ | 0.13  & 0.84 & 0.03
\cr
$\nspiii$ |
0.29 & 0.024 & $\sim 0$ | 0.028 & 0.038 & 0.27  & $\sim 0$ | 0.0  & 1.0 & 0.0
\cr
$\nspiv$ |
0.21 & 0.078 & 0.014    | 0.022 & 0.106 & 0.20  & $\sim 0$ | 0.0  & 1.0 & 0.0
\cr
$\nsmi$ |
0.25 & 0.084 & $\sim0$ | 0.004 & 0.056 & 0.27  & $\sim 0$ | 0.0   & 1.0  & 0.0
\cr
$\nsmii$ |
0.33 & 0.0   & $\sim0$ | 0.008 & 0.0   & 0.32  & $\sim 0$ | 0.14  & 0.73 & 0.13
\cr
$\nsmiv$ |
0.22 & 0.11 & $\sim 0$ | 0.004 & 0.085 & 0.25  & $\sim 0$ | 0.0  & 1.0 & 0.0
\cr
$\nsmv$ |
0.30 & 0.030 & $\sim 0$ | $\sim0$ & 0.022 & 0.31  & $\sim 0$ | 0.0  & 1.0 & 0.0
\cr
$\nsmvi$ |
0.32 & 0.016 & $\sim 0$ | 0.003   & 0.011 & 0.32  & $\sim 0$ | 0.0  & 1.0 & 0.0
\cr
$\nsmvii$ |
0.23 & 0.098 & 0.001    | 0.003   & 0.084 & 0.25  & $\sim 0$ | 0.0  & 1.0 & 0.0
\cr
$\nsmiii$ |
0.25 & 0.078 & $\sim0$ | $\sim0$ & 0.069 & 0.26  & $\sim 0$ | 0.0 & 1.0  & 0.0
\endtable
 \hrule \vskip .04in \hrule

 \bigskip
 \titlestyle{\tenpoint
 Table \brs d: A tabulation of some important branching ratios
for the $\sns$ scenarios delineated in Fig.~\boundaries d.
Results are quoted for a single $l$ or $\nu$ type (\eg\ $l=e$);
for the $\cntwo$ particle-antiparticle and antiparticle-particle
channels are summed together.
}
 \smallskip
\def\tstrut{\vrule height 12pt depth 4pt width 0pt}
 \thicksize=0pt
 \hrule \vskip .04in \hrule
\begintable
 \    | \multispan{2} \tstrut\hfil $\cpone\rta$ \hfil |
        \multispan{3} \tstrut\hfil $\cntwo\rta$ \hfil |
        \multispan{3} \tstrut\hfil $\slepl\rta$ \hfil |
        \multispan{2} \tstrut\hfil $\snu\rta$\hfil  \nr
 Scenario | $\snu l$ & $l$-3-body |
            $l\slepr$ & $\nu\snu$ & $l$-3-body |
            $\cpone\nu$ & $l \cnone$ & $l\cntwo$ |
            $\cpone l$ & $\cntwo\nu$ \cr
$\snspix$ |
0.35 & 0.02 | 0.16 & 0.16 & 0.006 |
0.48 & 0.23 & 0.29 |   0.0 & 0.0    \cr
$\snspvii$ |
0.0 & 0.21 | 0.14 & 0.0 & 0.08 |
0.31 & 0.53 & 0.16 |  0.036 & 0.009     \cr
$\snspviii$ |
0.006 & 0.076 | 0.20 & 0.0 & 0.10 |
0.25 & 0.62 & 0.13 |  0.0 & 0.0     \cr
$\snsmxii$ |
0.33 & $\sim0$ | 0.002 & 0.33 & $\sim 0$ |
0.37 & 0.36 & 0.27 |    0.0 & 0.0    \cr
$\snsmxiii$ |
0.33 & 0.004 | 0.002 & 0.33 & 0.004 |
0.05 & 0.92 & 0.03 |    0.0 & 0.0    \cr
$\snsmxiv$ |
0.33 & 0.001 | $\sim 0$ & 0.33 & 0.001 |
0.25 & 0.59 & 0.16 |  0.0 & 0.0     \cr
$\snsmxi$ |
0.33 & $\sim 0$ | $\sim 0$ & 0.33 & 0.001 |
0.29 & 0.51 & 0.20 |   0.0 & 0.0
\endtable
 \hrule \vskip .04in \hrule
 \vfill
\endinsert

\pageinsert
 \titlestyle{\tenpoint
 Table \prods a: Per cent of SUSY particles produced at Tevatron in
$2\rta 2$ subprocesses for $\dil$ scenarios. The quantity
$\wt\chi\wt\chi$ includes $\wt\chi_1^\pm\wt\chi_2^0$, while $\sq\,\sq$
doesn't include $\stopi\stopi$. The remaining sparticle production fraction is
taken up by associated production mechanisms.
}
 \smallskip
\def\tstrut{\vrule height 12pt depth 4pt width 0pt}
 \thicksize=0pt
 \hrule \vskip .04in \hrule
 \begintable
 Scenario | $\gl\,\gl$ | $\gl\,\sq$ | $\sq\,\sq$ | $\stopi\stopi$ |
$\wt\chi\wt\chi$
| $\wt\chi_1^\pm\wt\chi_2^0$ | $\slep\,\slep$ | $\slep\snu +\snu\snu$ \cr
$\dilpvii$ |  0.5  | 2.5 | 6.6 | 5.3 | 79 | 20 | 0.8 | 2.1 \nr
$\dilpii$ |  0.7  | 4.4  | 12  | 7.2 | 63 | 30 | 1.3 | 6.4 \nr
$\dilpvi$ |  0.8  | 4.1  | 13  | 6.8 | 59 | 32 | 3.7 | 7.1 \nr
$\dilpi$ |  0.3  | 1.5  | 6.3 | 7.6 | 74 | 35 | 1.5 | 4.6 \nr
$\dilpv$ |  0.07 | 0.3  | 2.9 | 5.9 | 78 | 43 | 3.1 | 6.9 \nr
$\dilpviii$ |  0.04 | 0.23 | 3.6 | 5.5 | 76 | 42 | 3.7 | 7.2 \nr
$\dilpiii$ |  0.01 | 0.05 | 1.2 | 3.7 | 82 | 46 | 4.4 | 7.2 \nr
$\dilpiv$ |  0.00 | 0.00 | 0.14| 2.0 | 85 | 47 | 6.0 | 6.2 \nr
$\dilmvii$ |  3.9 | 23  | 41  | 1.3 | 10 | 4.6| 1.9 | 10 \nr
$\dilmviii$ |  3.1 | 19  | 33 | 2.8 | 17 | 9.9 | 1.6 | 14 \nr
$\dilmii$ |  1.0 | 7.5 | 21  | 6.1 | 44 | 26 | 2.3 | 10 \nr
$\dilmix$ |  1.7 | 10  | 31  | 2.5 | 29 | 15 | 3.9 | 12 \nr
$\dilmvi$ |  1.0 | 5.4 | 16  | 7.0 | 49 | 29 | 4.2 | 9.9 \nr
$\dilmi$ |  0.7 | 4.3 | 18  | 4.4 | 46 | 26 | 4.3 | 14 \nr
$\dilmv$ |  0.1 | 0.6 | 5.6 | 3.8 | 66 | 38 | 6.4 | 12 \nr
$\dilmiii$ |  0.02| 0.08| 1.9 | 2.4 | 73 | 42 | 7.5 | 12 \nr
$\dilmiv$ |  0.00| 0.00| 0.3 | 0.3 | 74 | 38 | 12  | 13
\endtable
 \hrule \vskip .04in \hrule

\bigskip
 \titlestyle{\tenpoint
 Table \prods b: Per cent of SUSY particles produced at Tevatron in
$2\rta 2$ subprocesses for $\sdil$ scenarios. The quantity
$\wt\chi\wt\chi$ includes $\wt\chi_1^\pm\wt\chi_2^0$, while $\sq\,\sq$
doesn't include $\stopi\stopi$. The remaining sparticle production fraction is
taken up by associated production mechanisms.
}
 \smallskip
\def\tstrut{\vrule height 12pt depth 4pt width 0pt}
 \thicksize=0pt
 \hrule \vskip .04in \hrule
 \begintable
 Scenario | $\gl\,\gl$ | $\gl\,\sq$ | $\sq\,\sq$ | $\stopi\stopi$ |
$\wt\chi\wt\chi$
| $\wt\chi_1^\pm\wt\chi_2^0$ | $\slep\,\slep$ | $\slep\snu +\snu\snu$ \cr
$\sdilpiii$ |  0.03  | 0.1 | 0.8   | 4.7 | 88 | 53 | 1.4 | 2.5 \nr
$\sdilpvii$ |  0.02  | --- | 0.3   | 3.5 | 91 | 54 | 1.4 | 2.0 \nr
$\sdilpviii$ |  0.02  | 0.01 | 0.4 | 2.8 | 92 | 55 | 1.3 | 1.5 \nr
$\sdilmv$ |  0.09  | 0.1 | 0.8     | 3.7 | 87 | 53 | 2.0 | 3.1 \nr
$\sdilmviii$ |  0.02 | 0.05 | 0.5  | 1.8 | 90 | 52 | 2.5 | 3.1 \nr
$\sdilmix$ |  0.03 | 0.02 | 0.5    | 0.4 | 89 | 50 | 3.1 | 3.9
\endtable
 \hrule \vskip .04in \hrule
\vfill
\endinsert

\pageinsert
 \titlestyle{\tenpoint
 Table \prods c: Per cent of SUSY particles produced at Tevatron in
$2\rta 2$ subprocesses for $\ns$ scenarios. The quantity
$\wt\chi\wt\chi$ includes $\wt\chi_1^\pm\wt\chi_2^0$, while $\sq\,\sq$
doesn't include $\stopi\stopi$. The remaining sparticle production fraction is
taken up by associated production mechanisms.
}
 \smallskip
\def\tstrut{\vrule height 12pt depth 4pt width 0pt}
 \thicksize=0pt
 \hrule \vskip .04in \hrule
 \begintable
 Scenario | $\gl\,\gl$ | $\gl\,\sq$ | $\sq\,\sq$ | $\stopi\stopi$ |
$\wt\chi\wt\chi$
| $\wt\chi_1^\pm\wt\chi_2^0$ | $\slep\,\slep$ | $\slep\snu +\snu\snu$ \cr
$\nspi$ |  0.4  | 2.4  | 6.8 | 3.7 | 72 | 19 | 2.1 | 9.6 \nr
$\nspii$ |  0.4  | 2.3  | 9.5 | 3.4 | 47 | 23 | 3.7 | 30  \nr
$\nspviii$ |  0.1  | 0.4  | 3.6 | 2.6 | 57 | 31 | 11  | 22  \nr
$\nspv$ |  0.00 | 0.1  | 2.0 | 2.4 | 59 | 33 | 14  | 20  \nr
$\nspvii$ |  0.04 | 0.15 | 1.3 | 3.0 | 60 | 33 | 18  | 16  \nr
$\nspvi$ |  0.02 | 0.09 | 1.4 | 1.6 | 57 | 31 | 17  | 22  \nr
$\nspiii$ |  0.01 | 0.02 | 0.6 | 1.0 | 56 | 31 | 22  | 20  \nr
$\nspiv$ |  0.00 | 0.00 | 0.2 | 1.1 | 53 | 29 | 29  | 16  \nr
$\nsmi$ |  1.2  | 7.9  | 24  | 1.5 | 16 | 7.9| 7.8 | 34 \nr
$\nsmii$ |  0.5  | 3.0  | 13  | 2.8 | 30 | 18 | 5.3 | 41 \nr
$\nsmiv$ |  0.5  | 2.3  | 14  | 1.2 | 27 | 13 | 18  | 30 \nr
$\nsmv$ |  0.3  | 1.6  | 10  | 2.3 | 35 | 20 | 12  | 33 \nr
$\nsmvi$ |  0.1  | 0.9  | 7.2 | 2.3 | 42 | 24 | 13  | 31 \nr
$\nsmvii$ |  0.05 | 0.2  | 3.8 | 1.2 | 37 | 20 | 22  | 32 \nr
$\nsmiii$ |  0.01 | 0.1  | 2.4 | 1.2 | 42 | 24 | 22  | 30
\endtable
 \hrule \vskip .04in \hrule

\bigskip
 \titlestyle{\tenpoint
 Table \prods d: Per cent of SUSY particles produced at Tevatron in
$2\rta 2$ subprocesses for $\sns$ scenarios. The quantity
$\wt\chi\wt\chi$ includes $\wt\chi_1^\pm\wt\chi_2^0$, while $\sq\,\sq$
doesn't include $\stopi\stopi$. The remaining sparticle production fraction is
taken up by associated production mechanisms.
}
 \smallskip
\def\tstrut{\vrule height 12pt depth 4pt width 0pt}
 \thicksize=0pt
 \hrule \vskip .04in \hrule
 \begintable
 Scenario | $\gl\,\gl$ | $\gl\,\sq$ | $\sq\,\sq$ | $\stopi\stopi$ |
$\wt\chi\wt\chi$
| $\wt\chi_1^\pm\wt\chi_2^0$ | $\slep\,\slep$ | $\slep\snu +\snu\snu$ \cr
$\snspix$ |  0.2      | 0.6     | 3.1  | 3.7  | 72 | 42 | 5.7 | 11 \nr
$\snspvii$ |  $\sim 0$ | 0.01     | 0.03 | 1.2  | 77 | 44 | 12 | 8.6 \nr
$\snspviii$ |  $\sim 0$ | $\sim 0$ | 0.04 | 0.54 | 75 | 44 | 14 | 10 \nr
$\snsmxii$ |  1.4 | 6.4 | 14  | 3.8 | 43 | 27 | 3.5 | 20 \nr
$\snsmxiii$ |  1.4 | 8.6 | 18  | 2.6 | 35 | 19 | 6.0 | 19 \nr
$\snsmxiv$ |  1.0 | 6.4 | 15  | 3.3 | 41 | 25 | 4.8 | 20 \nr
$\snsmxi$ |  1.1 | 5.9 | 14  | 3.8 | 43 | 26 | 4.8 | 19
\endtable
 \hrule \vskip .04in \hrule
\vfill
\endinsert

\pageinsert
 \titlestyle{\tenpoint
 Table \xsecsbkgnd: Cross sections in fb after cuts given in the text, for
various background processes. The $W$ and $Z$ backgrounds include decays
to $\tau$ leptons. Non-parenthetical (parenthetical) numbers in the $3\ell$
case are cross sections for $0+1$ jets (any number of jets).
}
 \smallskip
\def\tstrut{\vrule height 12pt depth 4pt width 0pt}
 \thicksize=0pt
 \hrule \vskip .04in \hrule
 \begintable
 Process | $\etmiss$ | $1\ell$ | $OS$ | $SS$ | $3\ell$ | $4\ell$ \cr
$W+jets$ | 1450 | $1.1\times 10^6$ | --- | --- | --- | --- \nr
$Z+jets$ | 1065 | 6850        | --- | --- | --- | --- \nr
$t\bar t (178)$ | 200         | 491|0.02|0.35|0.07 (0.18) | --- \nr
$WW$     | 1.2  | 106         |31.7| --- | 0.01 (0.05) | --- \nr
$WZ$     | 0.1  | 3.9         |0.17| 1.8| 0.3 (0.4) | --- \nr
$total$ | 2716.3 | $1.1\times 10^6$|31.9| 2.15| 0.38 (0.63) | ---
\endtable
 \hrule \vskip .04in \hrule
\vfill
\endinsert

\pageinsert
 \titlestyle{\tenpoint
 Table \xsecs a: Cross sections in fb after cuts given in the text, for
various signals for the $\dil$ scenarios. Non-parenthetical (parenthetical)
numbers in the $3\ell$ case are cross sections for $0+1$ jets
(any number of jets).
}
 \smallskip
\def\tstrut{\vrule height 12pt depth 4pt width 0pt}
 \thicksize=0pt
 \hrule \vskip .04in \hrule
 \begintable
 Scenario | $\mgl$ | $\tanb$ | $\etmiss$ | $1\ell$ | $OS$ | $SS$ |
$3\ell$ | $3\ell / SS$ | $4\ell$ |  $5\ell$ \cr
$\dilpvii$ |282 |2.0 |479 |2010| 53.9 | 53.9 | 112 (168)   | 2.1 |32.4| 4.3\nr
$\dilpii$ |295 |9.0 |301 |1050| 16.2 | 18.6 | 16.2 (26.8) | 0.9 | ---  | ---
\nr
$\dilpvi$ |310 |15.0|304 |310 | 9.7  | 7.0  | 1.1 (3.2)   | 0.2 | ---  | ---
\nr
$\dilpi$ |346 |3.2 |103 |491 | 16.1 | 20.4 | 42.0 (56.2) | 2.1 |2.0 | 0.8 \nr
$\dilpv$ |431 |4.5 |23.5|128 | 8.8  | 4.8  | 16.0 (22.8) | 3.3 |0.8 | --- \nr
$\dilpviii$ |435 |8.0 |20.2|120| 5.5 | 3.2 | 4.7 (6.3)   | 1.5 |0.3 | --- \nr
$\dilpiii$ |503 |5.0 |9.2 |55.0| 2.8  | 3.0  | 4.2 (5.7)   | 1.4 |0.3 | 0.03
\nr
$\dilpiv$ |609 |2.0 |1.6 |20.6| 2.2  | 1.1  | 4.1 (5.6)  | 3.8 |0.2 | 0.03 \nr
$\dilmvii$ |232 |2.0 |1340|2530| 39.3 | 78.5 | 0.0 (5.8) | --- | ---  | --- \nr
$\dilmviii$ |243 |3.2 |1080 |2130 | 36.6 | 64.4 | 1.3 (10.1) | 0.02 |1.3| ---
\nr
$\dilmii$ |295 |9.0 |297 |483 | 6.7  | 6.3  | 1.0 (1.4)   | 0.2 | ---  | ---
\nr
$\dilmix$ |301 |2.2 |196 |480 | 17.3 | 5.9  | 5.4 (8.8)   | 0.9 |0.3 | 0.3 \nr
$\dilmvi$ |310 |15.0|278 |261 | 8.9  | 2.8  | 0.8 (1.2)   | 0.3 | ---  | ---
\nr
$\dilmi$ |346 |3.2 |72.8|207 | 6.1  | 1.2  | 1.2 (2.5)   | 1.0 | ---  | --- \nr
$\dilmv$ |431 |4.5 |12.4|30.3| 3.4  | 0.1  | 0.9 (1.1)   | 6.7 | ---  | --- \nr
$\dilmiii$ |503 |5.0 |5.0 |15.1| 3.5  | 0.04 | 0.5 (0.8)   | 13.0|0.04|0.02 \nr
$\dilmiv$ |609 |2.0 |4.0 |8.5 | 1.1  | 0.04 | 0.07 (0.08) | 1.9 | ---  | ---
\endtable
 \hrule \vskip .04in \hrule
\bigskip
 \titlestyle{\tenpoint
 Table \xsecs b: Cross sections in fb after cuts given in the text, for
various signals for the $\sdil$ scenarios. Non-parenthetical (parenthetical)
numbers in the $3\ell$ case are cross sections for $0+1$ jets
(any number of jets).
}
 \smallskip
\def\tstrut{\vrule height 12pt depth 4pt width 0pt}
 \thicksize=0pt
 \hrule \vskip .04in \hrule
 \begintable
 Scenario | $\mgl$ | $\tanb$ | $\etmiss$ | $1\ell$ | $OS$ | $SS$ |
$3\ell$ | $3\ell / SS$ | $4\ell$ |  $5\ell$ \cr
$\sdilpiii$  | 471 | 15.0| 15.4| 51.4| 2.3  | 1.1  | 4.1 (5.6)
| 3.6 |0.21| --- \nr
$\sdilpvii$  | 503 | 5.0 | 12.1| 40.0| 1.5  | 0.93 | 4.0 (5.1)
| 4.3 | 0.25 | 0.03 \nr
$\sdilpviii$ | 510 | 2.0 | 10.9| 38.9| 1.4  | 1.0  | 4.2 (5.2)
| 4.1 | 0.12 | 0.03 \nr
$\sdilmv$    | 471 | 15.0 | 11.2 | 45.4 | 2.3  | 1.5  | 4.9 (6.7)
| 3.2 | 0.24 | 0.03 \nr
$\sdilmviii$ | 503 | 5.0 | 7.7 | 27.4   | 1.6  | 0.7  | 3.9 (5.0)
| 5.5 | 0.16 | 0.05 \nr
$\sdilmix$   | 510 | 2.0 | 3.7 | 18.4   | 0.8  | 0.1  | 0.14 (0.28)
| 1.0 | 0.01 | ---
\endtable
 \hrule \vskip .04in \hrule
\vfill
\endinsert

\pageinsert
 \titlestyle{\tenpoint
 Table \xsecs c: Cross sections in fb after cuts given in the text, for
various signals for the $\ns$ scenarios.
Non-parenthetical (parenthetical) numbers in the $3\ell$
case are cross sections for $0+1$ jets (any number of jets).
}
 \smallskip
\def\tstrut{\vrule height 12pt depth 4pt width 0pt}
 \thicksize=0pt
 \hrule \vskip .04in \hrule
 \begintable
 Scenario | $\mgl$ | $\tanb$ | $\etmiss$ | $1\ell$ | $OS$ | $SS$ |
$3\ell$ |  $3\ell / SS$ | $4\ell$ |  $5\ell$ \cr
$\nspi$ |296 |2.2 |411 | 393 | 75.4 | 7.0  | 26.3 (43.9) | 3.8  | 5.3  | ---
\nr
$\nspii$ |324 |9.0 |141 | 983 | 51.1 | 17.4 | 24.6 (47.2) | 1.4  | 3.2
| --- \nr
$\nspviii$ |424 |8.0 |15.9| 214 | 8.5  | 2.3  | 7.7 (10.7)  | 3.4  | 0.34
| --- \nr
$\nspv$ |471 |4.5 |8.3 | 126 | 6.1  | 2.9  |5.4 (6.5)  | 1.9  | 0.07 | --- \nr
$\nspvii$ |491 |2.2 |7.9 | 96.5| 4.2  | 3.0  |3.3 (4.2)  | 1.1  | 0.06 | ---
\nr
$\nspvi$ |492 |7.0 |5.1 | 96.9| 5.7  | 1.3  |1.9 (2.5)  | 1.5  | 0.15 | --- \nr
$\nspiii$ |550 |5.0 |2.7 | 53.9| 3.1  | 2.6  | 1.2 (1.5) | 0.45 | 0.03 | ---
\nr
$\nspiv$ |603 |2.0 |1.9 | 31.2| 2.3  | 1.5  | 2.6 (3.4) | 1.7  | 0.05 | --- \nr
$\nsmi$ |296 |2.2 |257 | 809 | 22.7 | 22.2 | 14.1 (27.8) | 0.6  | --- |0.50 \nr
$\nsmii$ |324 |9.0 |119 | 784 | 19.2 | 8.4  | 2.8 (6.5) | 0.3  | 0.9  | --- \nr
$\nsmiv$ |368 |2.0 |38.3| 242 | 13.7 | 5.1  | 10.8 (18.2) | 2.1 | 0.4 | --- \nr
$\nsmv$ |373 |4.5 |34.6| 306 | 14.1 | 7.4  | 1.9 (4.4)   | 0.2 | --- | --- \nr
$\nsmvi$ |400 |7.0 |21.8| 239 | 13.9 | 4.5  | 0.5 (0.8)   | 0.1 | --- | --- \nr
$\nsmvii$ |450 |3.0 |6.5 | 101 | 7.6  | 1.9  | 6.4 (9.0) | 3.4 | 0.16 | --- \nr
$\nsmiii$ |477 |5.0 |4.2 | 84.5| 6.2  | 1.1  | 4.5 (5.8) | 4.0 | 0.2   | ---
\endtable
 \hrule \vskip .04in \hrule
\bigskip
 \titlestyle{\tenpoint
 Table \xsecs d: Cross sections in fb after cuts given in the text, for
various signals for the $\sns$ scenarios.
Non-parenthetical (parenthetical) numbers in the $3\ell$
case are cross sections for $0+1$ jets (any number of jets).
}
 \smallskip
\def\tstrut{\vrule height 12pt depth 4pt width 0pt}
 \thicksize=0pt
 \hrule \vskip .04in \hrule
 \begintable
 Scenario | $\mgl$ | $\tanb$ |
$\etmiss$ | $1\ell$ | $OS$ | $SS$ |
$3\ell$ | $3\ell / SS$ | $4\ell$ | $5\ell$ \cr
$\snspix$ | 390 | 8.0 |
31.6 | 55.5 | 23.8 | 0.81 | 4.6 (7.4) | 5.7 | 0.94 | --- \nr
$\snspvii$ | 570 | 2.2 |
3.63 | 30.1 | 3.2 | 0.96 | 6.3 (8.2) | 6.6 | 0.12 | --- \nr
$\snspviii$ | 591 | 4.5 |
5.8 | 16.5 | 2.2 | 0.59 | 3.1 (3.9) | 5.3 | 0.28 | 0.03 \nr
$\snsmxii$ | 290 | 9.0 |
207 | 864 | 30.9 | 10.1 | 1.5 (3.0) | 0.15 | ---  | --- \nr
$\snsmxiii$ | 300 | 3.0 |
157 | 581 | 21.4 | 8.1 | 1.8 (2.7) | 0.22 | ---  | --- \nr
$\snsmxiv$ | 300 | 5.0 |
153 | 635 | 23.8 | 6.0 | 0.36 (3.2) | 0.06 | --- | --- \nr
$\snsmxi$ | 301 | 6.0 |
138 | 627 | 23.0 | 5.1 | 1.1 (4.0) | 0.21 | ---  | ---
\endtable
 \hrule \vskip .04in \hrule
\vfill
\endinsert

\pageinsert
\vbox{\phantom{0}
\vskip 7.0in
\phantom{0}
\vskip .5in
\hskip -22pt
\special{ insert scr:boundariesdil.ps}
\vskip -1.45in }
\vfill
\centerline{\vbox{\hsize=12.4cm
\Tenpoint
\baselineskip=12pt
\noindent
Figure~\boundaries a,b:
We plot the boundaries of allowed \mgltb\
parameter space for the dilaton-equivalent and
string-scale-unified dilaton-equivalent scenarios
for both $\mu>0$ and $\mu<0$.  Values of $\rbtau$ for a given
$\tanb$ are given on the right-hand axis. The numbers indicate
the different cases considered for each scenario (see text). They are
positioned so that the actual \mgltb\ value associated with a point
is at the lower left-hand side of the number.
}}
\endinsert

\pageinsert
\vbox{\phantom{0}
\vskip 7.0in
\phantom{0}
\vskip .5in
\hskip -22pt
\special{ insert scr:boundariesns.ps}
\vskip -1.45in }
\vfill
\centerline{\vbox{\hsize=12.4cm
\Tenpoint
\baselineskip=12pt
\noindent
Figure~\boundaries c,d:
We plot the boundaries of allowed \mgltb\
parameter space for the minimal-supergravity and
string-scale-unified minimal-supergravity scenarios
for both $\mu>0$ and $\mu<0$.  Values of $\rbtau$ for a given
$\tanb$ are given on the right-hand axis. The numbers indicate
the different cases considered for each scenario (see text). They are
positioned so that the actual \mgltb\ value associated with a point
is at the lower left-hand side of the number.
}}
\endinsert

\pageinsert
\vbox{\phantom{0}
\vskip 7.0in
\phantom{0}
\vskip .5in
\hskip -22pt
\special{ insert scr:massesdil.ps}
\vskip -1.45in }
\vfill
\centerline{\vbox{\hsize=12.4cm
\Tenpoint
\baselineskip=12pt
\noindent
Figure~\masses a,b:
We plot the masses of the various superpartner particles
in terms of the ratio $m_i/\mgl$ ($i$ denoting a particular sparticle)
as a function of $\mgl$ for the $\dil$ and $\sdil$ models.
The scatter in the points indicates the
variation as $\tanb$ is allowed to vary at fixed $\mgl$.
The band corresponding to a given particle type $i$ is labelled,
according to its relative position at {\it large} $\mgl$, on the r.h.s.
of the graph.
}}
\endinsert

\pageinsert
\vbox{\phantom{0}
\vskip 7.0in
\phantom{0}
\vskip .5in
\hskip -22pt
\special{ insert scr:massesns.ps}
\vskip -1.45in }
\vfill
\centerline{\vbox{\hsize=12.4cm
\Tenpoint
\baselineskip=12pt
\noindent
Figure~\masses c,d:
We plot the masses of the various superpartner particles
in terms of the ratio $m_i/\mgl$ ($i$ denoting a particular sparticle)
as a function of $\mgl$ for the $\ns$ and $\sns$ models.
The scatter in the points indicates the
variation as $\tanb$ is allowed to vary at fixed $\mgl$.
The band corresponding to a given particle type $i$ is labelled,
according to its relative position at {\it large} $\mgl$, on the r.h.s.
of the graph.
}}
\endinsert

\pageinsert
\vbox{\phantom{0}
\vskip 8.5in
\phantom{0}
\vskip .5in
\hskip -60pt
\special{ insert scr:etmiss.ps}
\vskip -2.95in }
\vfill
\centerline{\vbox{\hsize=12.4cm
\Tenpoint
\baselineskip=12pt
\noindent
Figure~\etmissfig:
The $\etmiss$ cross section (in fb) for
$p\anti p$ collisions at $\sqrt s=1.8\tev$ is plotted for all
the numbered scenarios in Fig.~\boundaries\ as a function
of $\mgl$; $\dil$, $\sdil$, $\ns$ and $\sns$ cases are indicated
by $d$, $D$, $m$ and $M$, respectively. The $\mgl$ mass value
can be used to identify the scenario within each such case.
}}
\endinsert

\pageinsert
\vbox{\phantom{0}
\vskip 5.5in
\phantom{0}
\vskip .5in
\hskip -30pt
\special{ insert scr:ratesdil.ps}
\vskip -1.45in }
\vfill
\centerline{\vbox{\hsize=12.4cm
\Tenpoint
\baselineskip=12pt
\noindent
Figure~\rates a,b:
We display the same-sign dilepton ($SS$) and
tri-lepton ($3\ell$) cross sections (fb) for $\sqrt s=1.8\tev$ $p\anti p$
collisions for the $\dil$ and $\sdil$
scenarios defined in Fig.~\boundaries a,b.
No restrictions are placed on the number of associated jets.
The format is $[SS,3\ell]$ or $(SS,3\ell)$
where cases with square (curved) brackets have $\mhl<105\gev$ ($>105\gev$).
The dashed line indicates the approximate boundary
beyond which detection of a $SS$ or $3\ell$ signal would not be
possible for an integrated luminosity $L$ of $1\fbi$ or smaller.
The arrows, labelled by $\slepr$ and $\cpone$,
indicate the approximate $\mgl$ values corresponding to
$\mslepr=95\gev$ and $\mcpone=95\gev$, respectively.
The width of each arrow reflects the range
of $\mgl$ values obtained as $\tanb$ is varied at
fixed $\mslepr,\mcpone=95\gev$.
For $\mslepr<95\gev$, $\mcpone<95\gev$, $\mhl<105\gev$ detection of
$\slepr\slepr$, $\cpone\cmone$, $Z\hl$
production would be possible at LEP-200 with $L=500\pbi-1\fbi$.
}}
\endinsert

\pageinsert
\vbox{\phantom{0}
\vskip 5.5in
\phantom{0}
\vskip .5in
\hskip -30pt
\special{ insert scr:ratesns.ps}
\vskip -1.45in }
\vfill
\centerline{\vbox{\hsize=12.4cm
\Tenpoint
\baselineskip=12pt
\noindent
Figure~\rates c,d:
We display the same-sign dilepton ($SS$) and
tri-lepton ($3\ell$) cross sections (fb) for $\sqrt s=1.8\tev$ $p\anti p$
collisions for the $\ns$ and $\sns$
scenarios defined in Fig.~\boundaries c,d.
No restrictions are placed on the number of associated jets.
The format is $[SS,3\ell]$ or $(SS,3\ell)$
where cases with square (curved) brackets have $\mhl<105\gev$ ($>105\gev$).
The dashed line indicates the approximate boundary
beyond which detection of a $SS$ or $3\ell$ signal would not be
possible for an integrated luminosity $L$ of $1\fbi$ or smaller.
The arrows, labelled by $\slepr$ and $\cpone$,
indicate the approximate $\mgl$ values corresponding to
$\mslepr=95\gev$ and $\mcpone=95\gev$, respectively.
The width of each arrow reflects the range
of $\mgl$ values obtained as $\tanb$ is varied at
fixed $\mslepr,\mcpone=95\gev$.
For $\mslepr<95\gev$, $\mcpone<95\gev$, $\mhl<105\gev$ detection of
$\slepr\slepr$, $\cpone\cmone$, $Z\hl$
production would be possible at LEP-200 with $L=500\pbi-1\fbi$.
}}
\endinsert

\pageinsert
\vbox{\phantom{0}
\vskip 7.0in
\phantom{0}
\vskip .5in
\hskip -22pt
\special{ insert scr:lepdil.ps}
\vskip -1.45in }
\vfill
\centerline{\vbox{\hsize=12.4cm
\Tenpoint
\baselineskip=12pt
\noindent
Figure~\lep a,b:
We plot the contours for $\mslepr=83,95\gev$ (dots),
$\mcpone=83,95\gev$ (dashes) and $\mhl=81,105\gev$ (dotdash)
within  the allowed \mgltb\
parameter space for the dilaton-equivalent and for the
string-scale-unified dilaton-equivalent scenarios,
for both $\mu>0$ and $\mu<0$.
The lower and upper mass values for each particle
type represent rough upper limits for the mass reach at LEP-II
with $\sqrt s=176,200\gev$, respectively.
In the $\sdilp$ case, $\mslepr\geq 83\gev$ for all allowed $\mgl,\tanb$,
and only the $\mslepr=95\gev$ contour appears.
In the $\dilp$ and $\sdilp$ windows, the $\mhl=81\gev$ contour
is barely visible in the low-$\tanb$, low-$\mgl$ parameter space corner.
}}
\endinsert

\pageinsert
\vbox{\phantom{0}
\vskip 7.0in
\phantom{0}
\vskip .5in
\hskip -22pt
\special{ insert scr:lepns.ps}
\vskip -1.45in }
\vfill
\centerline{\vbox{\hsize=12.4cm
\Tenpoint
\baselineskip=12pt
\noindent
Figure~\lep c,d:
We plot the contours for $\mslepr=83,95\gev$ (dots),
$\mcpone=83,95\gev$ (dashes) and $\mhl=81,105\gev$ (dotdash)
within  the allowed \mgltb\
parameter space for the minimal-supergravity and
for the string-scale-unified minimal-supergravity scenarios,
for both $\mu>0$ and $\mu<0$.
The lower and upper mass values for each particle
type represent rough upper limits for the mass reach at LEP-II
with $\sqrt s=176,200\gev$, respectively.
In the $\nsm$ case, only the $\mslepr=83\gev$ contour (barely) appears
since $\mslepr\leq 95\gev$ throughout all of the allowed parameter space.
In the $\snsp$ window, the $\mhl=81\gev$ contour
is barely visible in the low-$\tanb$, low-$\mgl$ parameter space
corner. In the $\nsm$ model,
$\mhl\leq 105\gev$ for all allowed $\mgl,\tanb$ values,
and only the $\mhl=81\gev$ contour appears.
}}
\endinsert

\end